\documentclass[prd, onecolumn, nofootinbib, floatfix]{revtex4-1}
\pdfoutput=1

\usepackage{float}
\usepackage{hyperref}
\usepackage{amsmath}
\usepackage{graphicx}
\usepackage{dcolumn}
\usepackage{bm}
\usepackage{epsfig}
\usepackage{amssymb,latexsym,mathrsfs}
\usepackage{graphicx}
\usepackage{color}

\hypersetup{
    colorlinks=true,
    linkcolor=red,
    citecolor=blue,
    urlcolor=magenta,
} 

\newcommand{\be}{\begin{equation}}
\newcommand{\ee}{\end{equation}}
\newcommand{\bea}{\begin{eqnarray}}
\newcommand{\eea}{\end{eqnarray}}

\newcommand{\lcdm}{$\Lambda$CDM} 
\newcommand{\pps}{\Delta^2_{\mathcal R}} 
\newcommand{\ppsf}{\Delta^2_{{\mathcal R},0}} 
\newcommand{\mnu}{m_\nu}

\begin{document}

\title{Exploring Early and Late Cosmology with Next Generation Surveys} 

\author{Guilherme Brando${}^{*1,2}$, Eric V.\ Linder${}^{3,4}$} 
\affiliation{
${}^1$PPGCosmo, CCE - Universidade Federal do Esp\'irito Santo,
 29075-910, Vit\'oria, ES, Brazil\\ 
${}^2$Institute of Cosmology \& Gravitation, University of Portsmouth, Dennis Sciama Building, Burnaby Road, Portsmouth, PO1 3FX, United Kingdom\\ 
${}^3$Berkeley Center for Cosmological Physics \& Berkeley Lab, 
University of California, Berkeley, CA 94720, USA\\ 
${}^4$Energetic Cosmos Laboratory, Nazarbayev University, 
Nur-Sultan, Kazakhstan 010000 
}
\email{gbrando@cosmo-ufes.org}

\date{\today}

\begin{abstract}
Perturbations from inflation evolve into large scale structure of the 
late universe, and encode abundant cosmic structure formation physics. 
We allow freedom in the primordial power spectrum, rather than assuming a 
power law scale dependence, to study its impact on cosmological parameter 
determination. Combining various 
generations of cosmic microwave background (CMB) data and galaxy redshift 
survey data, we investigate the constraints on reconstruction of the 
primordial curvature perturbation power spectrum and the late time 
cosmology, especially the sum of neutrino masses. We quantify  how each 
successive generation, in CMB and galaxy surveys, provides significant 
improvements, often by factors of several. By  using CMB polarization 
information over a broad range of angular scales, and galaxy redshift 
data in many bins of redshift, one can allow inflationary freedom 
and still constrain parameters comparably to assuming power law dependence. 
The primordial power spectrum 
can be reconstructed at the subpercent level in a dozen wavenumber bins, 
while simultaneously fitting the sum of neutrino  masses to 14 meV.  
\end{abstract} 

\maketitle


\section{Introduction}

The vast array of large scale structure in the present universe formed 
from primordial seeds, possibly laid down as quantum fluctuations during 
early universe inflation. Cosmic structure is a convolution of the 
primordial perturbations and the evolution throughout cosmic history (also 
known as transfer functions). Thus it carries important information on 
both the early and late universe. 

Generally one assumes the form of the primordial perturbations is known, 
in terms of a power law (or power law plus running) scale dependence, 
and then measures structure through the cosmic microwave background (CMB) 
and galaxy surveys. When measuring structure at different cosmic times, 
one can extract the time dependence of its growth. This is influenced by 
the cosmic expansion history (and hence dark energy), the matter density 
(including dark matter), and neutrino mass (which also gives an additional 
scale dependence). 

These are all fundamental quantities of great import. However, one might 
wonder whether the assumptions about the primordial, i.e.\ initial, 
perturbation power spectrum could influence the results. This has been 
explored in various papers (e.g.\ 
\cite{deput14,kin00,haz13,zhao13,hann01,hlo11,gau12,dival16,gar15,rav16,can16,hunt15,haz18,haz19,sloz19,chluba15,beutl19,zeng18,domn19}). Another research area seeks to reconstruct the primordial power spectrum 
from data (see, e.g., \cite{handley19,lecl19,sobs,plc18inflation} 
and many others). Here we investigate both -- allowing inflationary 
freedom and examining its effect on late time cosmology, while seeing how 
late time data aids in reconstructing the primordial power spectrum. 
In  particular, CMB data probes the density power spectrum on many scales, 
while galaxy redshift data can probe it at many  times; together they can  
deconvolve the evolution from the initial conditions. 

We emphasize taking a function-free form (which we will refer to as free form) primordial curvature perturbation power spectrum, 
to allow for features of various sorts (including a power law) from inflation  
physics. While there are models that predict steps, bends, or oscillations  
linear or logarithmic in scale (see, e.g., 
\cite{sloz19,obied17,zeng18,domn19,lhui17,haz18,chen16,haz16} 
and references therein), we work with a free form that would allow the data 
itself to reconstruct its preferred behavior. 
While no finite number of free variables gives complete freedom, this does give far more freedom than fixing to power law, and previous literature (e.g.\  \cite{deput14}) has established that 20 bins give sufficient freedom for cases of most interest.
Of course one would also compare
the data to particular compelling models, but without clear guidance the
phenomenological approach we take here can provide useful first tests. 

Our goal is to consider how the next generation of CMB experiments and galaxy 
surveys will improve the simultaneous fitting of early and late time 
cosmology. We look at CMB experiments (present and future) alone, and 
in combination with next generations of galaxy redshift surveys, with  
particular focus on reconstructing the primordial power spectrum and the 
late time constraint on the sum of neutrino masses.  

In Sec.~\ref{sec:method} we lay out the method for implementing 
inflationary freedom, treating the galaxy power spectrum, and analyzing the 
constraints on inflation, cosmology, and astrophysical parameters. 
Section~\ref{sec:experiments} summarizes the surveys we 
will consider, of various generations, and the types and range of data 
used. In  
Sec.~\ref{sec:results} we discuss the results  of various combinations 
of data, by probe and generation, and summarize the prospects for reconstruction 
of the inflationary power spectrum and constraint on the sum of neutrino masses. 
We conclude in Sec.~\ref{sec:concl}.

\section{Inflationary Freedom and Cosmology} \label{sec:method} 

The growth of large scale structure can be viewed as a convolution of 
the initial conditions, in terms of the primordial curvature perturbation power spectrum 
$\pps$, with a scale and time dependent transfer function $T(k,z)$ 
describing the evolution: 
\be 
P_k^\delta(z)=\pps(k) * T(k,z)\ , 
\ee 
where $\mathcal{R}$ is the curvature perturbation and $\delta$ is the 
fractional matter density perturbation, with $k$ the Fourier mode and 
$z$ the redshift. For the remainder of the article we will write 
$P_k^\delta$ simply as $P_k$. 

From this form we can immediately see that data $P_k(z)$ at several 
redshifts allows the possibility of separating the primordial contribution 
from the evolution, and measuring both. Using the CMB at the redshift of 
last scattering $z_{\rm lss}\approx 1090$ and galaxy surveys in several 
redshift slices $z\lesssim5$ in principle permits simultaneous 
constraints on the early and late universe. In particular, 
future galaxy surveys that extend to $z\gtrsim2$ and have multiple 
redshift bins should have large numbers of linear modes -- even larger 
than the number of modes mapped by the CMB -- hence with greater 
statistical leverage and easier to interpret, and a range of redshifts 
to separate out the several cosmological influences. 

The scale dependence in the transfer function generally comes from 
when perturbation modes enter the horizon (i.e.\ radiation vs matter 
dominated epoch), which is well accounted for in Boltzmann codes, 
from nonlinear mode coupling (which we will avoid by restricting to the 
linear density perturbation regime), and from scale dependent cosmology such 
as the influence of neutrino mass (which we include). 
The time dependence involves the cosmic background expansion, and hence 
the matter density and dark energy properties. We will work within the 
flat $\Lambda$CDM model to focus on the matter density, neutrino mass, 
and mass fluctuation amplitude $\sigma_8$. 

Conventionally one assumes a particular form for $\pps(k)$, most usually 
a power law in $k$, sometimes with another parameter describing the 
running of the power law index. That is, 
\be 
\ppsf(k)=\ppsf(k_{\rm piv})\,\left(\frac{k}{k_{\rm piv}}\right)^{n_s-1+(\alpha_s/2)\ln(k/k_{\rm piv})}\ , 
\ee 
where $k_{\rm piv}$ is a pivot wavenumber, $n_s$ the power law index, and 
$\alpha_s$ the running. 
While the simplest theories of inflation predict such a form, interest 
has increased in forms that could include oscillations or other 
features. Therefore we want to allow great inflationary freedom. 
Since there is particular interest recently in oscillations, we avoid 
adopting any form that might give spurious oscillations for ringing, 
such as splines, principal components, or wavelets. 

For the primordial power spectrum (PPS) we use 20 bins in $k$, giving a 
highly model independent form capable of reconstructing oscillations or 
features. Thus the PPS becomes 
\be 
\pps(k)=\ppsf(k)\,\left[1+\sum h_i\,B_i(k)\right]\ , \label{eq:ppsh} 
\ee 
where $B_i$ is the top hat function (1 inside the bin centered at $k_i$, 
0 outside), and $h_i$ is the free parameter giving the height of the 
deviation from the fiducial power law. That is, we allow the power law 
PPS to be multiplied in each bin by a factor $1+h_i$, and sample $h_i$ 
with a uniform prior in $[-0.8,0.8]$. 
We set $\ppsf$ by adopting the 
fixed Planck 2018 \cite{planck2018} values of $A_{s}=2.1 \times 10^{-9}$  
(which determines $\ppsf(k_{\rm piv})$ according to our conventions), 
$n_s=0.966$, and $\alpha_s=0$ for $k_{\rm piv}=0.05\,{\rm Mpc}^{-1}$. 
The bins are spaced logarithmically in $k$, from $10^{-3}$--$10^{-0.5}$ 
Mpc$^{-1}$, $2.5$ decades with 8 bins per decade so 20 bins total. 
To recap: the $h_i$ give the amplitude in each bin, and if they ``work together''  
they can choose an overall slope. The data determines this, the 
form Eq.~(\ref{eq:ppsh}) does not impose it. 

The matter density power spectrum $P_m$ is determined from the primordial 
curvature power spectrum through solution of the evolution equations 
within an appropriately modified version of the CLASS Boltzmann code \cite{class,jesus} 
(specifically, we generalize the initial conditions by introducing our binned PPS version in the code via the {\tt external\_Pk} 
mode feature and add a galaxy clustering likelihood module for the future data we use). We convert this to the 
galaxy power spectrum by including galaxy bias and redshift space 
distortions, as well as a nuisance contribution from nonlinear density 
effects and deviations from 
linear galaxy bias. We write the galaxy power spectrum 
$P_g$ as 
\be 
P_g(k,\mu,z)=\left[b(z)+f(z)\mu^2\right]^2 P_m(k,z)+P_0(z)\ , 
\ee 
where $b(z)$ is the linear bias factor, $f(z)$ is  the matter growth 
rate, $\mu$ is the cosine of the angle of the Fourier mode $\vec k$ 
with respect to the line of sight, and $P_0$ is a nuisance function  
allowing for power 
coming from misestimation of bias, nonlinearities, and shot noise. This 
form follows that used in analyses of BOSS galaxy redshift survey 
data \cite{cyr13,deput14}. 
Recall that at $z\gtrsim 2$, where we will focus, modes stay linear to 
larger $k$, so the range $k<10^{-0.5}\,{\rm Mpc}^{-1}$ used should be 
reasonably described by linear theory. 

The growth rate $f(z)$ will be determined by the background cosmological 
parameters (since we are here only using data at $z\gtrsim2$ we 
use flat $\Lambda$CDM), and we include parameters $b_j$ and 
$S_j$ for each redshift 
bin of the survey, where $S_j\equiv n_j P_0(z_j)$. For $b_j$ we take 
uniform priors in $[b_{j,{\rm fid}}-1.5, b_{j,{\rm fid}}+1.5]$, where the fiducial 
value depends on the survey (see the next section). For $S_j$ we take 
uniform priors in $[-0.5,0.5]$ for the additional, nuisance power. 
(Note $S_j=n_j P_0$ is from the unaccounted-for extra nuisance power, 
and is not the $nP_g$ of the survey.) 
The cosmological 
parameters are the dimensionless physical baryon density $\Omega_b h^2$, 
dimensionless physical  
cold dark matter density $\Omega_{cdm} h^2$, Hubble constant $H_0$ (written 
in units of km$\,s^{-1}{\rm Mpc}^{-1}$), 
sum of neutrino masses $\sum m_\nu$, 
and optical depth $\tau$. Recall that $A_s$ and $n_s$ are replaced as 
free parameters by the twenty $h_i$. Derived parameters may include the 
total matter density in units of the critical density, $\Omega_m$, and 
the mass fluctuation amplitude $\sigma_8$.

\section{Next Generation Surveys} \label{sec:experiments} 

One main focus is to explore how next generation surveys, of both the 
CMB and large scale structure clustering, will improve our knowledge 
of the primordial power spectrum, without assuming a functional form for it, i.e.\ allowing inflationary freedom. 
We consider two CMB surveys: the current Planck 2018 data and a future 
CMB-S4 experiment \cite{cmbs4sci,cmbs4des}. 
For galaxy redshift surveys, we consider two variations of the 
proposed next generation MegaMapper survey, called the Ideal and Fiducial  
versions \cite{MM2,MM1}. 

The Planck data is used through the Planck 2018 likelihood code available at MontePython 
\cite{montepython1,montepython2}. We use the Planck18 baseline likelihood: Planck TTTEEE, Planck low-T, low-E, and lensing. 

For the CMB-S4 experiment we create a mock data set assuming Planck 2018 
bestfit \lcdm\ parameter values ($\Omega_{b}h^{2} = 0.02238$, $\Omega_{cdm}h^{2} = 0.1201$, $H_{0} = 67.32$, $\tau_{reio} = 0.0543$, $n_{s}=0.966$, $A_{s}=2.1 \times 10^{-9}$, $\sum\mnu=0.06\,$eV), and we use the mock likelihood class for this future survey 
introduced in \cite{spreng18} (note that CMB-S4 experiment characteristics 
may have changed since then). Note that for CMB-S4 we do not add multipoles 
$\ell<30$ outside its range 
(i.e.\ do not also use Planck18 data), which, together with using a realization of the \lcdm\ model rather than measured data, can shift the best fit parameters relative to Planck18 data fits. 

MegaMapper mock data likelihood was created through a 
galaxy clustering likelihood module added to MontePython 
through a class module {\tt montepython/likelihoods\_class.py}, using 
the survey characteristics 
described in  
\cite{MM2} including galaxy bias $b_{j,{\rm fid}}$ and galaxy number 
density $n_j$ entering the shot noise added to the data. 
Both Ideal and Fiducial cases span the range $z=2$--5, though 
with a higher galaxy number density and finer redshift binning in the Ideal case (seven 
and four bins respectively, each bin with parameters for galaxy bias $b_j$ and 
modeling $S_j$). 

Thus we have 20 inflation parameters, 5 cosmology parameters (i.e.\ the PPS amplitude $A_s$ and power law index $n_s$ are replaced by the 20 inflation parameters), and 
8 or 14 astrophysical nuisance parameters, in addition to CMB nuisance 
parameters. 

Our scientific focus here is mainly on two aspects of the physics: the 
primordial power spectrum and the sum of the neutrino masses, though we will discuss 
the other parameter constraints, especially any unusual ones. We will analyze 
cases of data combinations in both a ``probe'' -- CMB only or CMB plus galaxy 
redshift survey -- and a ``generational'' sense -- current data, moderately future 
data (including MegaMapper Fiducial data), and further future data (including 
CMB-S4 or MegaMapper Ideal).

\section{Results} \label{sec:results}

\subsection{CMB only} \label{sec:cmbonly} 

\begin{figure}[htbp!] 
\centering
\includegraphics[width=0.85\columnwidth]{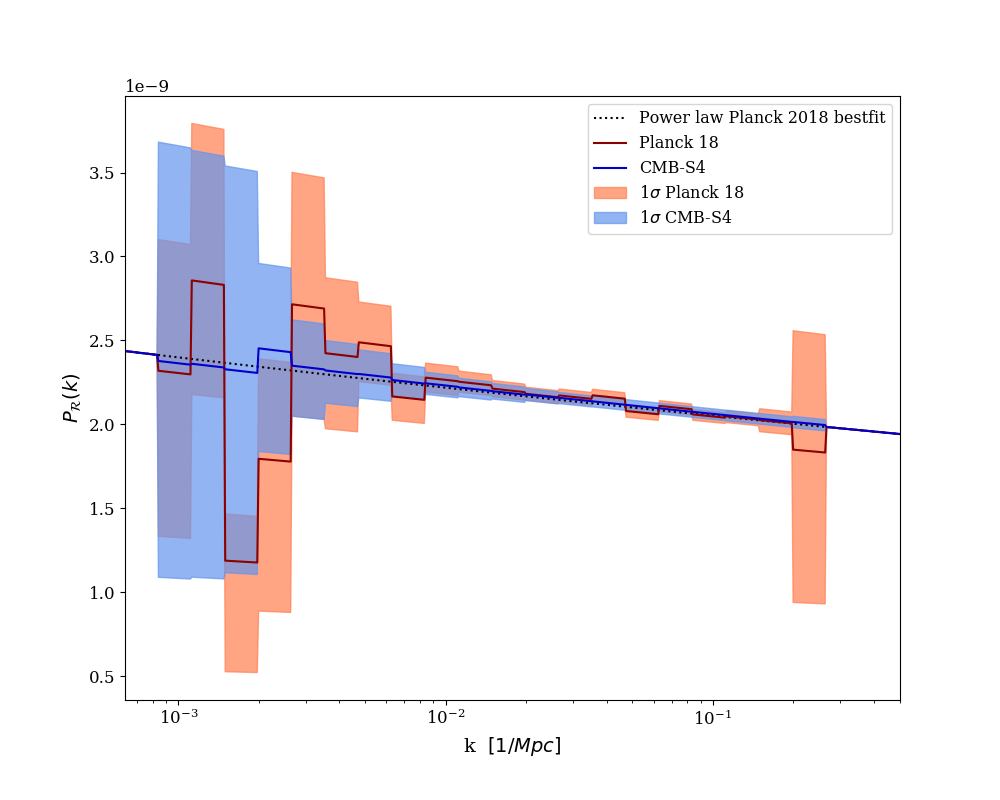}
\caption{Reconstructed primordial power spectrum using data from Planck18, with the 
mean reconstruction shown by the red line and 68\% CL uncertainty band in orange, 
and projected for CMB-S4 around the best fit Planck18 \lcdm\ cosmology (dotted black line), 
with the mean reconstruction shown by the blue line and 68\% CL uncertainty band 
in blue. Note the CMB-S4 mock data does not include data for $\ell<30$, giving larger 
uncertainties at the smallest $k$. 
} 
\label{fig:pl18s4recon}
\end{figure}

We begin with just the CMB data, and the current data from the Planck 2018 data set,  
which includes the lowest multipoles, important for constraining the small $k$ modes 
of the primordial power spectrum.  The PPS reconstruction is shown in 
Fig.~\ref{fig:pl18s4recon}. 
In general this agrees well with a power law, with the slope  
$n_s$ consistent with the usual power law Planck results. The significant, well-known 
deviation is in the third free bin of amplitude, $h_3$ at $k\in[1.78,2.37]\times 10^{-3}\,{\rm Mpc}^{-1}$. 
This is lower than the power law expectation at more than 95\% confidence level (CL), and 
corresponds to the dip in the temperature power spectrum data around $\ell\approx20$. Since there is freedom in the other $k$ bins, they 
rise slightly in amplitude to compensate, but still follow a very 
similar power law dependence.

\begin{figure}[htbp!] 
\centering
\includegraphics[width=0.95\columnwidth]{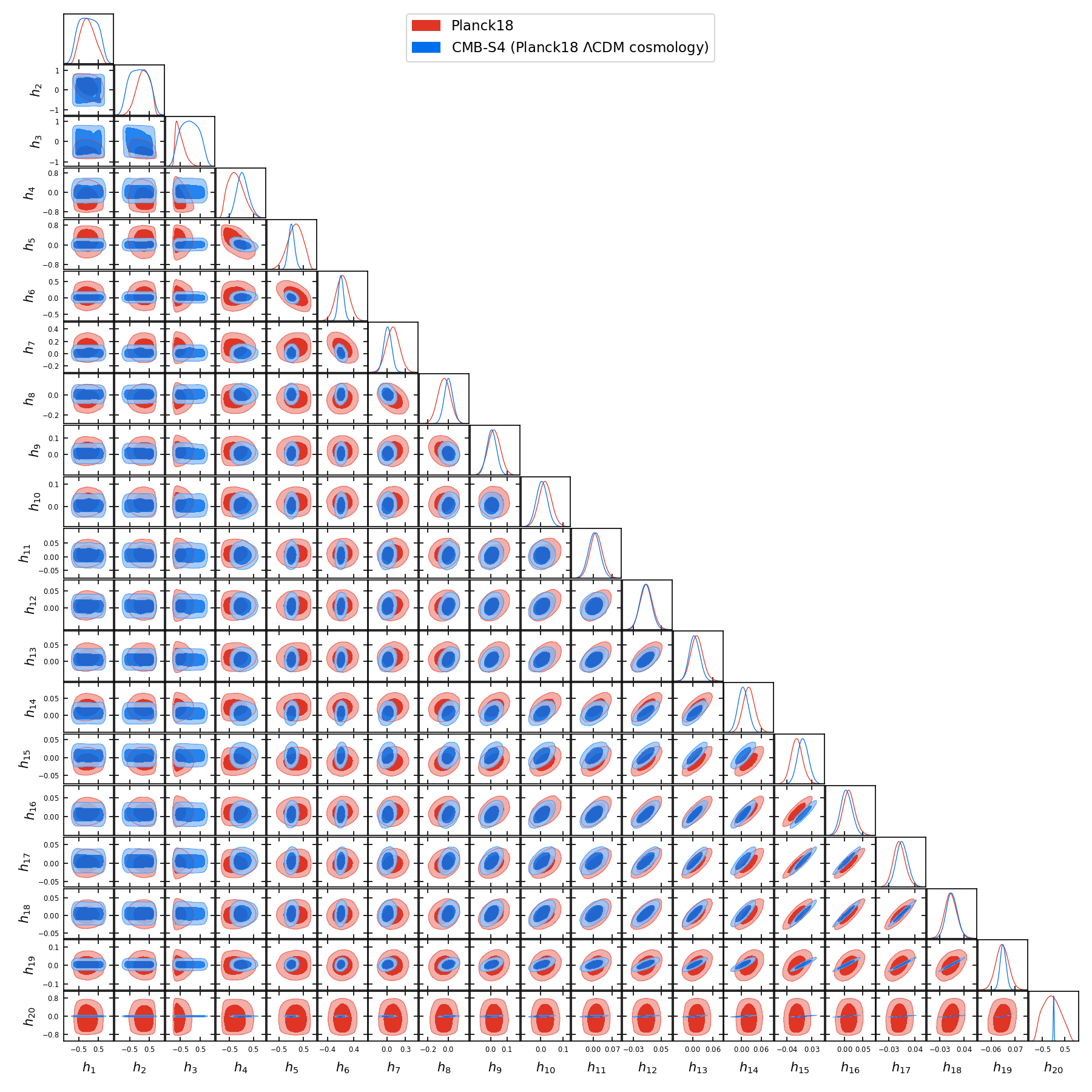} 
\caption{
Corner plot of constraints on the primordial power spectrum amplitudes $h_i$ in 20 bins of wavenumber $k$  
plotted using data from Planck18 (red), 
and projected for CMB-S4 (blue). Contours are 
at 68.3\% and 95.4\% confidence levels, and 
the marginalized 1D probability distributions 
are shown as well. 
} 
\label{fig:pl18s4h}
\end{figure}

The 20 PPS bin amplitude parameters $h_i$ are given in Table~\ref{tab:pl18s4h} and 
plotted in Fig.~\ref{fig:pl18s4h}. (All plots of posteriors 
in this work were generated using the Python package GetDist~\cite{getdist}.) 
Except for $h_3$ they are all 
consistent with zero (i.e.\ the standard power law PPS) at 95\% 
confidence level. The 68\% CL constraints range from uncertainties 
of $\sim1.5\%$ in $1+h_i$ (see Eq.~\ref{eq:ppsh}) at the tightest to 
$\sim50\%$ at the smallest $k$ (where cosmic variance is large) and 
largest $k$ (where Planck measurement precision begins to weaken).

\begin{table}[H]
    \centering
\begin{tabular}{|l||c|c|c|c||c|c|c|c|}
 \hline
Param & best-fit & mean$\pm\sigma$ & 95\% lower & 95\% upper & best-fit & mean$\pm\sigma$ & 95\% lower & 95\% upper \\ \hline

\rule{0pt}{1.1\normalbaselineskip}$100~\Omega_{b}h^{2}$ &$2.228$ & $2.222_{-0.02}^{+0.019}$ & $2.183$ & $2.26$ & $2.237$ & $2.238_{-0.0041}^{+0.004}$ & $2.229$ & $2.246$ \\
\rule{0pt}{1.1\normalbaselineskip}$\Omega_{cdm}h^{2}$ &$0.1225$ & $0.1222_{-0.0015}^{+0.0015}$ & $0.1191$ & $0.1253$ & $0.1205$ & $0.1202_{-0.00093}^{+0.00084}$ & $0.1184$ & $0.122$ \\
\rule{0pt}{1.1\normalbaselineskip}$H_0$ &$66.59$ & $65.52_{-1.2}^{+1.9}$ & $62.38$ & $68.26$ & $67.04$ & $67.11_{-0.57}^{+0.79}$ & $65.78$ & $68.31$ \\
\rule{0pt}{1.1\normalbaselineskip}$\tau_{reio }$ &$0.04342$ & $0.05441_{-0.0084}^{+0.0074}$ & $0.03828$ & $0.07122$ & $0.05387$ & $0.05692_{-0.0081}^{+0.008}$ & $0.041$ & $0.07327$ \\
\rule{0pt}{1.1\normalbaselineskip}$\sum m_{\nu}$ &$0.04126$ & $0.1747_{-0.17}^{+0.044}$ & $8.011\times10^{-6}$ & $0.4507$ & $0.07551$ & $0.07976_{-0.064}^{+0.038}$ & $2.101\times10^{-6}$ & $0.1637$ \\
\rule{0pt}{1.1\normalbaselineskip}$\sigma_{8}$ &$0.8037$ & $0.7908_{-0.027}^{+0.036}$ & $0.7259$ & $0.8512$ & $0.8084$ & $0.8087_{-0.0067}^{+0.0091}$ & $0.7935$ & $0.8226$\\
\rule{0pt}{1.1\normalbaselineskip}$\Omega_{m }$ &$0.3276$ & $0.3418_{-0.015}^{+0.027}$ & $0.305$ & $0.3857$ & $0.3197$ & $0.3186_{-0.0075}^{+0.011}$ & $0.303$ & $0.3359$ \\
\rule{0pt}{1.1\normalbaselineskip}$h_1$ &$-0.3354$ & $-0.03881_{-0.41}^{+0.34}$ & $-0.6922$ & $0.6651$ &$0.07525$ & $-0.0108_{-0.45}^{+0.45}$ & $-0.758$ & $0.754$ \\
\rule{0pt}{1.1\normalbaselineskip}$h_2$ &$0.3522$ & $0.1958_{-0.3}^{+0.39}$ & $-0.3862$ & $0.8$ &$-0.02108$ & $-0.0141_{-0.45}^{+0.45}$ & $-0.761$ & $0.752$ \\
\rule{0pt}{1.1\normalbaselineskip}$h_3$ &$-0.5113$ & $-0.4977_{-0.3}^{+0.076}$ & $-0.8$ & $-0.02429$ &$0.2763$ & $-0.02202_{-0.44}^{+0.44}$ & $-0.755$ & $0.748$ \\
\rule{0pt}{1.1\normalbaselineskip}$h_4$ &$-0.2626$ & $-0.2339_{-0.4}^{+0.26}$ & $-0.8$ & $0.3532$ &$-0.02197$ & $0.04807_{-0.28}^{+0.22}$ & $-0.424$ & $0.5605$ \\
\rule{0pt}{1.1\normalbaselineskip}$h_5$ &$0.2272$ & $0.1701_{-0.29}^{+0.35}$ & $-0.4029$ & $0.7672$ &$0.0342$ & $0.01304_{-0.14}^{+0.12}$ & $-0.2403$ & $0.2758$ \\
\rule{0pt}{1.1\normalbaselineskip}$h_6$ &$0.0392$ & $0.05489_{-0.2}^{+0.2}$ & $-0.3455$ & $0.4553$ &$-0.03092$ & $0.01094_{-0.089}^{+0.08}$ & $-0.157$ & $0.1831$ \\
\rule{0pt}{1.1\normalbaselineskip}$h_7$ &$0.03037$ & $0.09374_{-0.11}^{+0.11}$ & $-0.1193$ & $0.3067$ &$-0.02318$ & $0.00861_{-0.066}^{+0.061}$ & $-0.1185$ & $0.1364$ \\
\rule{0pt}{1.1\normalbaselineskip}$h_8$ &$-0.04971$ & $-0.03859_{-0.065}^{+0.063}$ & $-0.1667$ & $0.08929$ &$-0.01839$ & $0.005279_{-0.046}^{+0.044}$ & $-0.08373$ & $0.097$ \\
\rule{0pt}{1.1\normalbaselineskip}$h_9$ &$-0.0146$ & $0.02106_{-0.04}^{+0.039}$ & $-0.05855$ & $0.1006$ &$0.009427$ & $0.005752_{-0.031}^{+0.031}$ & $-0.0553$ & $0.06685$ \\
\rule{0pt}{1.1\normalbaselineskip}$h_{10}$ &$0.005821$ & $0.01999_{-0.03}^{+0.029}$ & $-0.03853$ & $0.07943$ &$-0.00317$ & $0.004445_{-0.027}^{+0.025}$ & $-0.04639$ & $0.05584$ \\
\rule{0pt}{1.1\normalbaselineskip}$h_{11}$ &$-0.0152$ & $0.01126_{-0.023}^{+0.022}$ & $-0.03422$ & $0.05717$ &$-0.01732$ & $0.003982_{-0.022}^{+0.022}$ & $-0.0401$ & $0.04701$ \\
\rule{0pt}{1.1\normalbaselineskip}$h_{12}$ &$-0.008292$ & $0.006973_{-0.019}^{+0.018}$ & $-0.03015$ & $0.04536$ &$0.007063$ & $0.005671_{-0.017}^{+0.017}$ & $-0.02803$ & $0.03938$ \\
\rule{0pt}{1.1\normalbaselineskip}$h_{13}$ &$-0.01717$ & $0.01232_{-0.019}^{+0.018}$ & $-0.02518$ & $0.04983$ &$0.004557$ & $0.006046_{-0.017}^{+0.016}$ & $-0.02522$ & $0.03754$ \\
\rule{0pt}{1.1\normalbaselineskip}$h_{14}$ &$0.001468$ & $0.02269_{-0.019}^{+0.018}$ & $-0.01407$ & $0.05999$ &$-0.004285$ & $0.005611_{-0.017}^{+0.015}$ & $-0.02527$ & $0.03766$ \\
\rule{0pt}{1.1\normalbaselineskip}$h_{15}$ &$-0.03524$ & $-0.01177_{-0.018}^{+0.016}$ & $-0.04632$ & $0.02338$ &$0.001419$ & $0.005827_{-0.016}^{+0.016}$ & $-0.02549$ & $0.03774$ \\
\rule{0pt}{1.1\normalbaselineskip}$h_{16}$ &$-0.008964$ & $0.01256_{-0.018}^{+0.016}$ & $-0.02156$ & $0.04747$ &$-0.0004784$ & $0.005597_{-0.016}^{+0.015}$ & $-0.0256$ & $0.03694$ \\
\rule{0pt}{1.1\normalbaselineskip}$h_{17}$ &$-0.02743$ & $-0.002141_{-0.018}^{+0.015}$ & $-0.0354$ & $0.0323$ &$-0.0004859$ & $0.005503_{-0.016}^{+0.015}$ & $-0.02561$ & $0.03713$ \\
\rule{0pt}{1.1\normalbaselineskip}$h_{18}$ &$-0.02281$ & $0.002186_{-0.019}^{+0.017}$ & $-0.03433$ & $0.03931$ &$-0.0007732$ & $0.005426_{-0.016}^{+0.016}$ & $-0.02637$ & $0.0374$ \\
\rule{0pt}{1.1\normalbaselineskip}$h_{19}$ &$-0.02507$ & $0.0008731_{-0.035}^{+0.034}$ & $-0.06837$ & $0.07003$ &$-0.0007021$ & $0.005276_{-0.017}^{+0.016}$ & $-0.0266$ & $0.03778$ \\
\rule{0pt}{1.1\normalbaselineskip}$h_{20}$ &$-0.2821$ & $-0.07702_{-0.46}^{+0.37}$ & $-0.8$ & $0.5931$ &$-0.0007574$ & $0.005613_{-0.018}^{+0.017}$ & $-0.02914$ & $0.04098$ \\
\hline
 \end{tabular} \\
\caption{Cosmological parameters and the primordial power spectrum amplitudes $h_i$ in the 20 bins in wavenumber $k$, 
for Planck18 data (left columns) and for CMB-S4 mock data generated from the Planck18 \lcdm\ cosmology (right columns). 
Param gives the parameter, and 95\% lower and upper 
give the 95.4\% confidence level lower and upper 
bounds on the parameter. Nuisance parameters not shown are marginalized over. 
} 
\label{tab:pl18s4h} 
\end{table}

While there is some covariance between the $h_i$, choosing 20 bins 
gives a happy medium where features -- if they exist -- could be well 
resolved but covariances are not excessive. This will become especially true when we combine CMB and galaxy data. 
We see no significant evidence for oscillatory features (e.g.\ low then high 
values of $h_i$) or steps, other than the well known dip around 
$k\approx2\times10^{-3}\,{\rm Mpc}^{-1}$. 

Moving to the projections for the future CMB-S4 experiment, shown 
also in  Figs.~\ref{fig:pl18s4recon} and \ref{fig:pl18s4h}  and 
the right hand set of columns of Table~\ref{tab:pl18s4h}, the 
PPS is properly consistent with the input power law mock data, and 
we see significant improvements to the PPS amplitude uncertainties. 
Beyond the third bin (i.e.\ $\ell\gtrsim30$) CMB-S4 demonstrates 
significant improvements over Planck, all the way to the highest $k$ bin, 
due to its better resolution and noise levels. Half of 
the amplitudes $1+h_i$ are determined to better than $\sim$2\%. This will 
make CMB-S4 a powerful probe of the primordial power spectrum and 
any features from the inflation physics. (Note we do not include 
any Planck data in the CMB-S4 case, so there is no $\ell<30$ data 
and for the first three $k$ bins the constraints are weaker 
than Planck.) 

The cosmology parameters (recall this is for free PPS) mostly follow the usual behaviors from power law PPS. Slight 
shifts in parameter values between Planck18 and CMB-S4 data generated from Planck18 cosmology 
can occur since Planck18 is data while CMB-S4 is mock data 
realized from a \lcdm\ input cosmology, and since our CMB-S4 realization lacks 
$\ell<30$ data (just as Planck18 and Planck18 minus low $\ell$ data 
could give slightly different parameter fits). 
These shifts
are found to be negligible, at most less than the 0.4 sigma level, and so not 
statistically significant. 
The cosmological parameter uncertainties with just Planck18 are comparable to those from the standard power law case, with the main difference being a factor of roughly two increase in uncertainties on $\sum\mnu$, $H_0$, and $\sigma_8$ due to the extra freedom in the PPS amplitude with scale.

\begin{figure}[htbp!] 
\centering
\includegraphics[width=0.95\columnwidth]{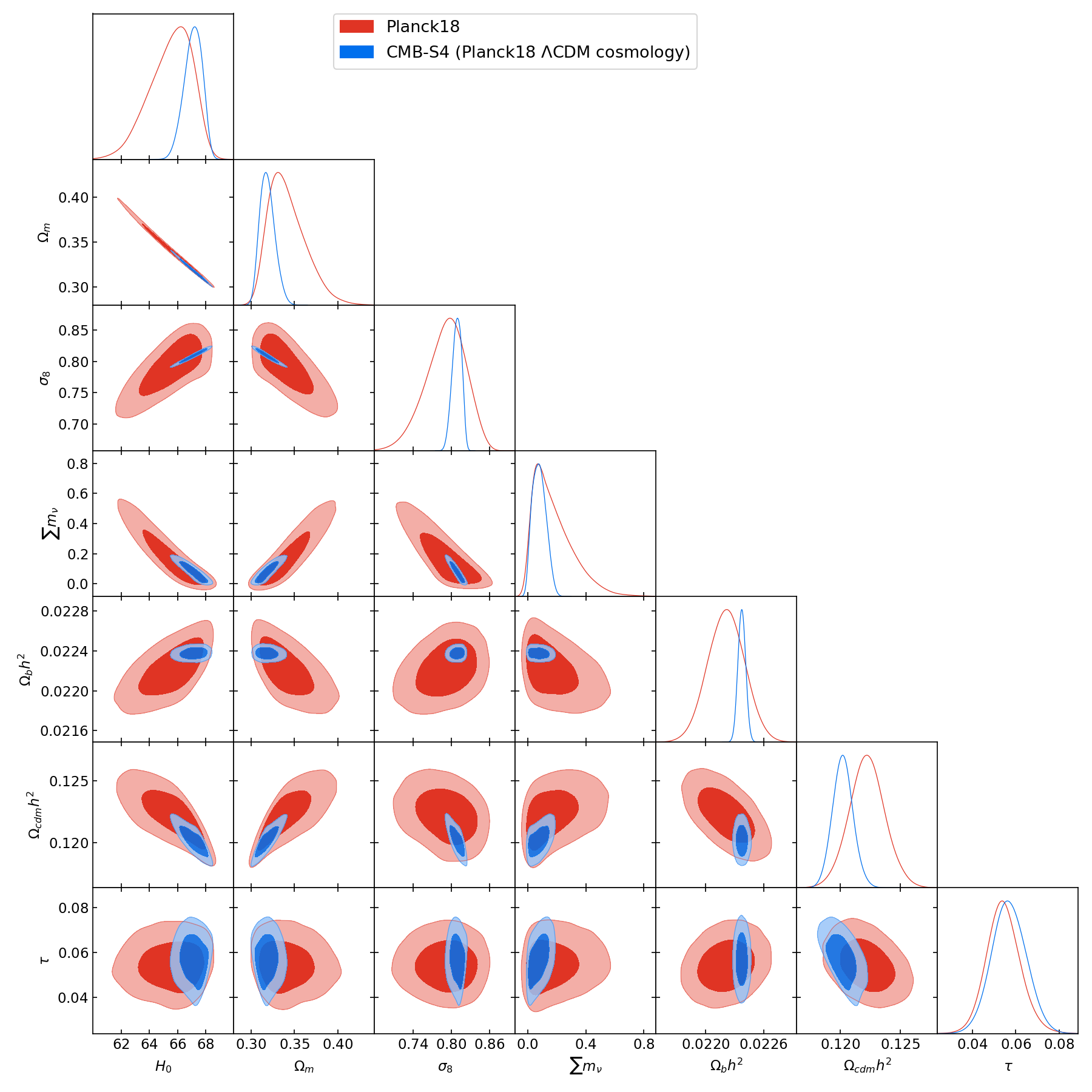}
\caption{
Corner plot of constraints on the cosmology parameters, including derived ones, are 
plotted using data from Planck18 (red), 
and projected for CMB-S4 (blue). CMB-S4 represents a 
consider tightening of cosmological constraints.  
} 
\label{fig:pl18s4cos}
\end{figure}

The more important aspect is the improvement in parameter uncertainties with future data. 
Articles such as \cite{cmbs4sci,cmbs4des} have shown the great 
advances in cosmological parameter determination enabled by CMB-S4; 
we find this holds when using free PPS as well. 
Confidence contours for the cosmology parameters are 
shown in Fig.~\ref{fig:pl18s4cos}, making clear the improvement. 
Since neutrino mass is one 
of the key aspects of this paper, we note that CMB-S4 reduces the 
uncertainty on the sum of the neutrino masses from a 95\% CL upper limit of  
$\sum\mnu\lesssim0.45\,$eV to 0.16\,eV. In the ``growth cosmology'' plane 
of $\sum\mnu$--$\sigma_8$ the reduction in uncertainty contour 
area is more than a factor of 10. We discuss neutrino mass constraints in greater 
detail in  Sec.~\ref{sec:summnu}.

\subsection{CMB plus Galaxy Surveys} \label{sec:cmbgal} 

The combination of the CMB as a probe of the primordial universe 
with the capability of galaxy surveys to deliver a tomographic 
view of the evolution of structure formation is especially powerful 
when the PPS is allowed to be free rather than locked into power 
law scale dependence. 

We now consider the far future case of a high precision CMB 
experiment (CMB-S4) with a high precision galaxy survey capable of 
superb tomography over $z=2$--$5$ (MegaMapper Ideal). This gives 
enhanced sensitivity to scale dependent physics, such as the PPS 
and neutrino mass. Results are shown in Table~\ref{tab:mmh}. The amplitudes $1+h_i$ now approach the 0.2\% 
level or even better, especially at higher $k$ due to the excellent deconvolution 
of the transfer function, i.e.\ evolution, from the primordial 
power spectrum. Only at the largest scales, $k\lesssim0.015$/Mpc, 
does the uncertainty exceed 1\%. Such precision will provide deep 
insight into any features in the inflationary physics. 
The covariances between $h_i$ are shown in Appendix~\ref{sec:apxbS}. 
(Note that for the CMB-S4 + MegaMapper Ideal case the neutrino mass 
has fluctuated a bit below that for CMB-S4 only 
in Table~\ref{tab:pl18s4h}, meaning less suppression and so the $h_i$ 
tend to be slightly negative though not by more than $1\sigma$.)

\begin{table}[H]
    \centering
\begin{tabular}{|l||c|c|c|c||c|c|c|c|}
\hline
Param & best-fit & mean$\pm\sigma$ & 95\% lower & 95\% upper & best-fit & mean$\pm\sigma$ & 95\% lower & 95\% upper \\ \hline

\rule{0pt}{1.1\normalbaselineskip}$100~\Omega_{b}h^{2}$ &$2.245$ & $2.241_{-0.013}^{+0.013}$ & $2.215$ & $2.268$ & $2.225$ & $2.225_{-0.0028}^{+0.0028}$ & $2.219$ & $2.23$ \\
\rule{0pt}{1.1\normalbaselineskip}$\Omega_{cdm}h^{2}$ &$0.1196$ & $0.12_{-0.0009}^{+0.0005}$ & $0.1188$ & $0.1215$ & $0.1194$ & $0.1196_{-0.00034}^{+0.00032}$ & $0.119$ & $0.1202$ \\
\rule{0pt}{1.1\normalbaselineskip}$H_0$ &$67.97$ & $67.56_{-0.38}^{+0.79}$ & $66.37$ & $68.54$ & $67.71$ & $67.54_{-0.25}^{+0.25}$ & $67.05$ & $68.02$ \\
\rule{0pt}{1.1\normalbaselineskip}$\tau_{reio }$ &$0.05391$ & $0.05413_{-0.0033}^{+0.0032}$ & $0.04765$ & $0.06073$ & $0.05497$ & $0.05471_{-0.00091}^{+0.00091}$ & $0.05292$ & $0.0565$ \\
\rule{0pt}{1.1\normalbaselineskip}$\sum m_{\nu}$ &$0.01977$ & $0.04514_{-0.045}^{+0.013}$ & $8.615\times 10^{-7}$ & $0.09908$ &$0.0368$ & $0.04844_{-0.013}^{+0.014}$ & $0.02158$ & $0.07562$ \\
\rule{0pt}{1.1\normalbaselineskip}$\sigma_{8}$ &$0.8187$ & $0.8146_{-0.003}^{+0.0045}$ & $0.8074$ & $0.8212$ & $0.8148$ & $0.8123_{-0.0023}^{+0.0022}$ & $0.8079$ & $0.8168$ \\
\rule{0pt}{1.1\normalbaselineskip}$\Omega_{m }$ &$0.3081$ & $0.3133_{-0.0043}^{+0.01}$ & $0.3015$ & $0.3282$ & $0.31$ & $0.3123_{-0.0033}^{+0.0034}$ & $0.3062$ & $0.3187$ \\
\rule{0pt}{1.1\normalbaselineskip}$h_1$ &$-0.0598$ & $-0.07303_{-0.38}^{+0.31}$ & $-0.7289$ & $0.5878$ &$0.4595$ & $0.006577_{-0.46}^{+0.46}$ & $-0.7009$ & $0.7636$ \\
\rule{0pt}{1.1\normalbaselineskip}$h_2$ &$0.02348$ & $0.01665_{-0.28}^{+0.27}$ & $-0.5167$ & $0.5634$ &$-0.1442$ & $-0.001408_{-0.39}^{+0.37}$ & $-0.6671$ & $0.6684$ \\
\rule{0pt}{1.1\normalbaselineskip}$h_3$ &$-0.4163$ & $-0.3033_{-0.23}^{+0.2}$ & $-0.7177$ & $0.109$ &$-0.04378$ & $-0.005165_{-0.22}^{+0.21}$ & $-0.4426$ & $0.4387$ \\
\rule{0pt}{1.1\normalbaselineskip}$h_4$ &$-0.1129$ & $-0.0614_{-0.15}^{+0.14}$ & $-0.3462$ & $0.2266$ &$-0.07495$ & $-0.003495_{-0.13}^{+0.12}$ & $-0.2522$ & $0.2465$ \\
\rule{0pt}{1.1\normalbaselineskip}$h_5$ &$-0.06181$ & $0.0138_{-0.092}^{+0.092}$ & $-0.1704$ & $0.199$ &$-0.04954$ & $-0.003597_{-0.071}^{+0.069}$ & $-0.1413$ & $0.1378$ \\
\rule{0pt}{1.1\normalbaselineskip}$h_6$ &$0.002678$ & $0.01869_{-0.057}^{+0.055}$ & $-0.09206$ & $0.1325$ &$-0.02603$ & $-0.006101_{-0.044}^{+0.043}$ & $-0.09166$ & $0.07955$ \\
\rule{0pt}{1.1\normalbaselineskip}$h_7$ &$0.007194$ & $0.007343_{-0.034}^{+0.035}$ & $-0.0618$ & $0.07659$ &$0.001492$ & $-0.005252_{-0.029}^{+0.029}$ & $-0.06131$ & $0.05167$ \\
\rule{0pt}{1.1\normalbaselineskip}$h_8$ &$0.001145$ & $-0.007705_{-0.03}^{+0.029}$ & $-0.06656$ & $0.05261$ &$0.004995$ & $-0.006739_{-0.024}^{+0.023}$ & $-0.05306$ & $0.04054$ \\
\rule{0pt}{1.1\normalbaselineskip}$h_9$ &$0.004989$ & $0.008361_{-0.032}^{+0.031}$ & $-0.05434$ & $0.07157$ &$-0.02778$ & $-0.001105_{-0.026}^{+0.026}$ & $-0.05245$ & $0.05036$ \\
\rule{0pt}{1.1\normalbaselineskip}$h_{10}$ &$0.001783$ & $0.001536_{-0.015}^{+0.014}$ & $-0.02684$ & $0.03112$ &$-0.003038$ & $-0.00514_{-0.011}^{+0.01}$ & $-0.02642$ & $0.01609$ \\
\rule{0pt}{1.1\normalbaselineskip}$h_{11}$ &$0.001936$ & $-0.0004983_{-0.013}^{+0.011}$ & $-0.02376$ & $0.0246$ &$-0.004895$ & $-0.003653_{-0.0081}^{+0.0084}$ & $-0.02024$ & $0.01292$ \\
\rule{0pt}{1.1\normalbaselineskip}$h_{12}$ &$-0.003505$ & $-0.001884_{-0.011}^{+0.0088}$ & $-0.02081$ & $0.01824$ &$-0.002216$ & $-0.002861_{-0.0056}^{+0.0057}$ & $-0.01436$ & $0.008479$ \\
\rule{0pt}{1.1\normalbaselineskip}$h_{13}$ &$-0.004738$ & $0.0003305_{-0.0092}^{+0.0071}$ & $-0.01509$ & $0.01701$ &$-0.001704$ & $-0.002775_{-0.0041}^{+0.0041}$ & $-0.01113$ & $0.005437$ \\
\rule{0pt}{1.1\normalbaselineskip}$h_{14}$ &$-0.0003906$ & $0.0005026_{-0.0084}^{+0.0064}$ & $-0.0135$ & $0.01575$ &$-0.002986$ & $-0.003812_{-0.0034}^{+0.0034}$ & $-0.01053$ & $0.002928$ \\
\rule{0pt}{1.1\normalbaselineskip}$h_{15}$ &$-0.002715$ & $-0.001394_{-0.0073}^{+0.0054}$ & $-0.01344$ & $0.01167$ &$-0.0031$ & $-0.002622_{-0.0027}^{+0.0026}$ & $-0.007802$ & $0.002705$ \\
\rule{0pt}{1.1\normalbaselineskip}$h_{16}$ &$-0.001334$ & $0.000247_{-0.0049}^{+0.0041}$ & $-0.008609$ & $0.009371$ &$-0.001665$ & $-0.002109_{-0.0021}^{+0.0021}$ & $-0.006296$ & $0.002054$ \\
\rule{0pt}{1.1\normalbaselineskip}$h_{17}$ &$-0.002041$ & $-0.0002224_{-0.0048}^{+0.0038}$ & $-0.008547$ & $0.008522$ &$-0.0006202$ & $-0.001137_{-0.0019}^{+0.0019}$ & $-0.004865$ & $0.002604$ \\
\rule{0pt}{1.1\normalbaselineskip}$h_{18}$ &$-0.0008104$ & $0.0002547_{-0.0041}^{+0.0035}$ & $-0.007308$ & $0.00795$ &$-0.001353$ & $-0.001856_{-0.0018}^{+0.0017}$ & $-0.005318$ & $0.001608$ \\
\rule{0pt}{1.1\normalbaselineskip}$h_{19}$ &$-0.0001978$ & $0.0002558_{-0.0029}^{+0.0026}$ & $-0.005191$ & $0.005824$ &$-0.000699$ & $-0.001001_{-0.0014}^{+0.0014}$ & $-0.003771$ & $0.001864$ \\
\rule{0pt}{1.1\normalbaselineskip}$h_{20}$ &$-0.0008644$ & $0.00001845_{-0.0015}^{+0.0015}$ & $-0.003015$ & $0.003055$ &$0.0007476$ & $0.0001666_{-0.001}^{+0.001}$ & $-0.00184$ & $0.002188$ \\
 \hline
 \end{tabular} \\
\caption{
Cosmological parameters and the primordial power spectrum amplitudes $h_i$ in the 20 bins in wavenumber $k$, 
for Planck18 plus MegaMapper Fiducial mock data (left columns) and for CMB-S4 plus MegaMapper Ideal mock data generated from the best fit Planck18 \lcdm\ cosmology (right columns). 
} 
\label{tab:mmh}
\end{table}

\begin{figure}[htbp!] 
\centering
\includegraphics[width=0.95\columnwidth]{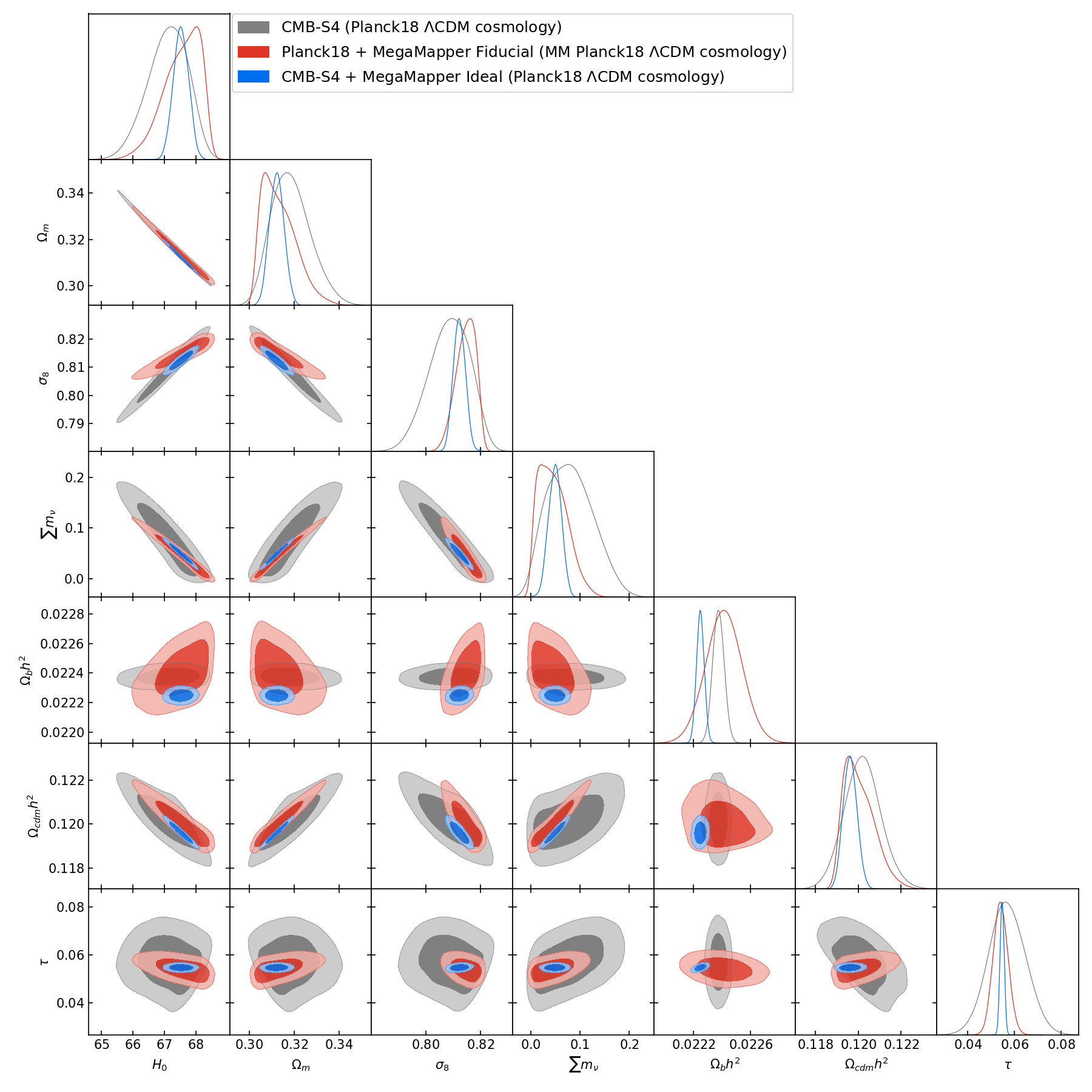}
\caption{
Corner plot of constraints on the cosmology parameters, included derived ones, are 
plotted using data from CMB-S4 (grey; as in Fig.~\ref{fig:pl18s4cos}), Planck18 plus MegaMapper Fiducial (red), and CMB-S4 plus 
MegaMapper Ideal (blue). 
} 
\label{fig:mmcos}
\end{figure}

Neutrino mass constraints improve to the $\sigma(\sum\mnu)\approx 0.014\,$eV level, allowing 
a clear detection of the minimal mass sum according to neutrino 
oscillation experiments 
and potentially significant distinction between the normal and inverted 
mass hierarchies. 
Most other cosmological parameters show similar 
improvements in precision, of order a factor 3--4. 
The cosmology parameter constraints are shown 
in Fig.~\ref{fig:mmcos}. The shift in $\Omega_b h^2$ upon 
adding galaxy data can be 
traced back to interaction between the free PPS and the nonlinear correction 
parameters $S_i$ (and the knock-on effects on other parameters).  
The astrophysical 
parameters $b_i$ and $S_i$ are also well constrained (see Appendix~\ref{sec:apxbS}).

\begin{figure}[htbp!] 
\centering
\includegraphics[width=0.85\columnwidth]{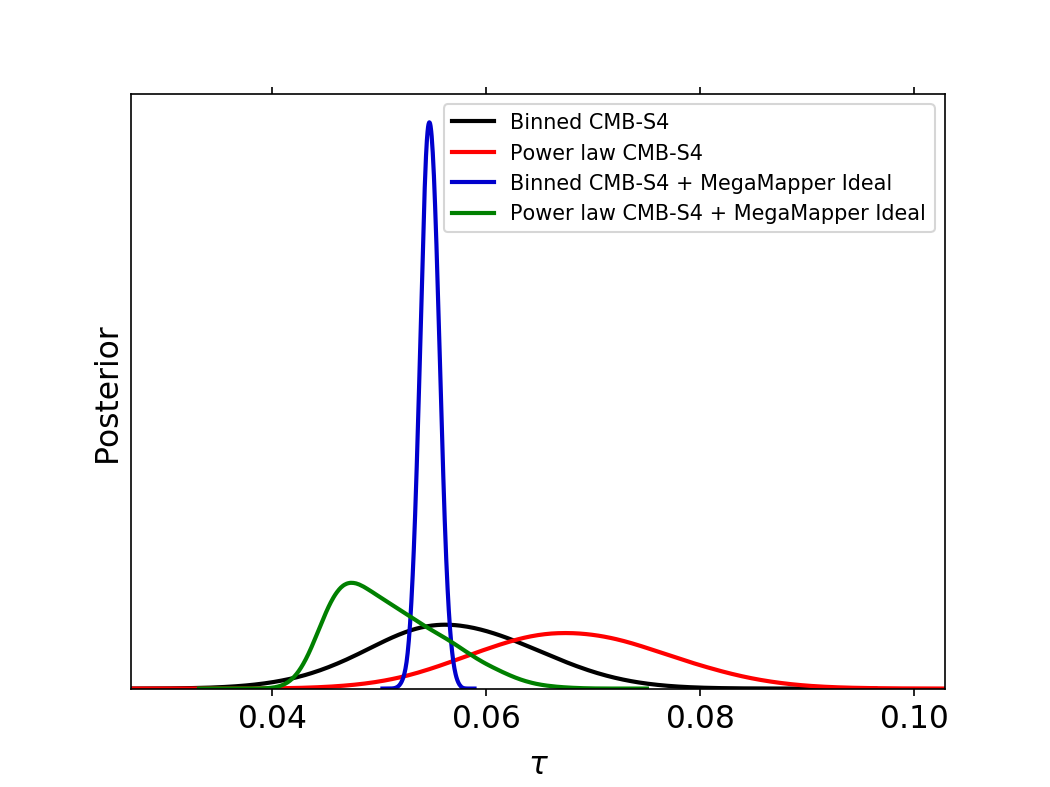}
\caption{
1D PDFs of the reionization optical depth $\tau$ are contrasted 
for the free, binned PPS and the standard restriction to power law PPS, 
for the cases of CMB-S4 alone and CMB-S4 plus MegaMapper Ideal. 
As discussed in the text, the highly peaked and tightly distributed form of the posterior for the binned version is a manifestation of the decoupling of short and large scales due to the free amplitudes in $k$ bins, which can be taken advantage of 
by the multiple redshift bins in the galaxy data to deconvolve the primordial conditions from  the transfer function. The same behavior is not present in the power law PPS, as we have only one degree of freedom when constraining the amplitude, i.e.\ shifting the PPS up or down as a whole. 
} 
\label{fig:tau_1d}
\end{figure}

Perhaps most unusual is the effect on the reionization optical depth 
$\tau$. 
In the standard power law PPS case the optical depth is 
degenerate 
with the amplitude of the primordial curvature perturbation power $A_s$ in  the form $A_s e^{-2\tau}$. 
In our free binned version of the PPS, due to the freedom allowed by the 20 amplitude bins, 
we can independently probe large and small scales of the PPS. 
In combining CMB and tomographic galaxy clustering one achieves excellent $\sigma(h_{i})$ at small scales, and these 
are translated into tight constraints on $\sigma(\tau)$; by contrast, in the power law case 
we cannot separate large from small scales in the PPS amplitude. 
This holds as well with CMB only data, where constraints
on $\tau$ match well projections in the literature, and even for Planck18
plus MegaMapper Fiducial (though we start to see some improvement there).  
The separation of scales becomes 
especially effective at the high redshifts of MegaMapper 
where smaller scales stay more linear, and most so for MegaMapper 
Ideal with its higher number density, plus CMB-S4 with its better 
resolution and noise. Figure~\ref{fig:tau_1d}
illustrates the 1D posterior distributions of the reionization optical depth in the
CMB-S4 + MegaMapper Ideal case (and also CMB-S4 alone) for the power law and free binned PPS cases. 
Comparing the blue curve with the black demonstrates the role of 
future galaxy clustering data, while keeping the same binned PPS model. 
Comparing the 
blue curve with the green -- i.e.\ fixing to the CMB+galaxy data sets -- demonstrates that using binned PPS instead 
of power law PPS is what gives the improved constraint on tau, as
discussed above.

In the nearer future is the case using the current CMB data from 
Planck18 and projections from the less ambitious MegaMapper Fiducial. 
This delivers quite respectable constraints on the PPS, a factor of a few better on $h_i$ at the higher $k$ than without galaxy data, though about a 
factor 2 worse than the CMB-S4 + MegaMapper Ideal case. The cosmology constraints for this case are also shown 
in Table~\ref{tab:mmh} and Fig.~\ref{fig:mmcos}. 
Again we see that in the ``growth'' plane of $\sum\mnu$--$\sigma_8$, the combination of CMB-S4 +  MegaMapper Ideal improves by a further factor of 10 
relative to CMB-S4 alone, and by a factor of several 
on Planck18 + MegaMapper Fiducial. 
We compare the full cosmology parameter results between 
the free PPS and power law PPS for CMB-S4 + MegaMapper Ideal 
and for Planck18 + MegaMapper Fiducial in Appendix~\ref{sec:apxpowerlaw}.

\subsection{Summary of Primordial Power Spectrum Constraints} 
\label{sec:sumpps} 

For inflation physics we see that the upcoming generation of CMB and large scale structure  
experiments can deliver strong improvements over the current 
constraints on the primordial curvature power spectrum. This will be 
highly useful in exploring for properties in inflation such as bends 
or steps in the inflationary potential or additional fields that would 
have signatures in oscillations or sharp features in the primordial 
power spectrum. Freedom beyond a power law scale dependence would 
be an important discovery and clue to high energy physics. 

\begin{figure}[htbp!] 
\centering
\includegraphics[width=0.85\columnwidth]{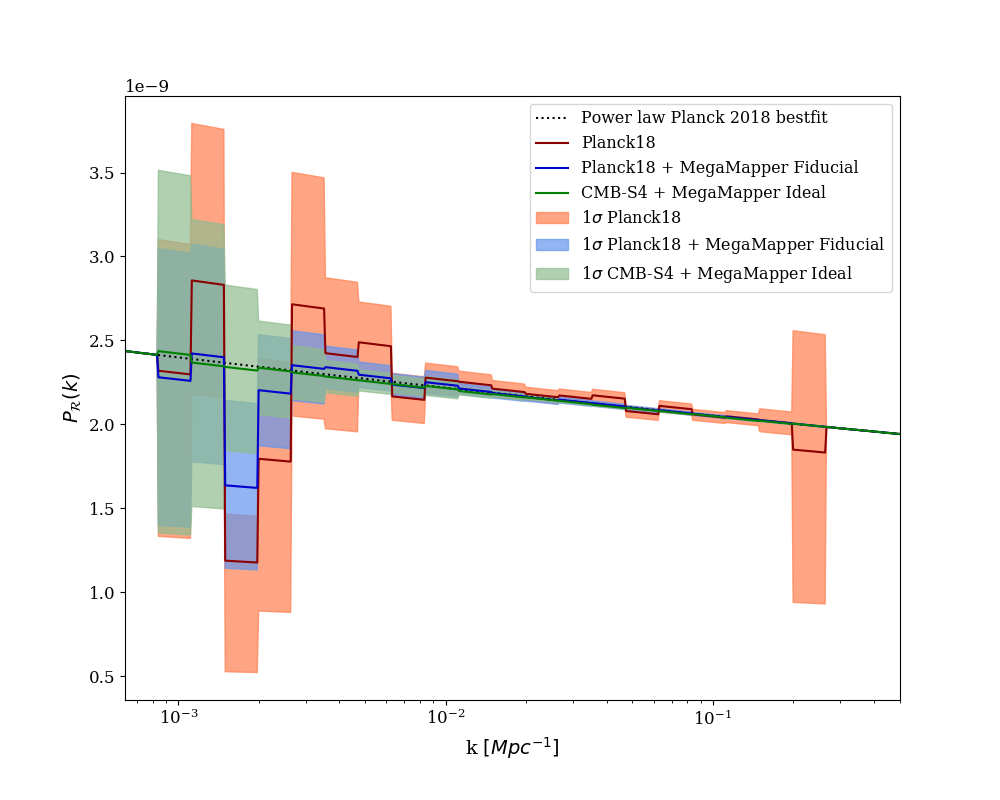}
\caption{
Reconstructed primordial power spectrum using actual data from Planck18 
(red, as in Fig.~\ref{fig:pl18s4recon}), Planck18 plus MegaMapper 
Fiducial (blue), and CMB-S4 plus MegaMapper Ideal (green). Future data 
will be able to significantly narrow in on, or find deviations from, 
power law behavior in scale.  
} 
\label{fig:sumpps}
\end{figure}

Figure~\ref{fig:sumpps} shows the constraints enabled by the next 
generation of experiments. Note this does not account for improvements 
in the low multipoles $\ell<30$; polarization data from 
satellite missions such as LiteBIRD \cite{litebird1,litebird2} (and possibly 
ground based experiments with sufficient systematics control) could 
lead to further gains there. Note that large scale structure data 
plays a significant role for $k\gtrsim{\rm few}\times 10^{-3}\,{\rm Mpc}^{-1}$  
even with excellent CMB data.

\begin{figure}[htbp!] 
\centering
\includegraphics[width=0.85\columnwidth]{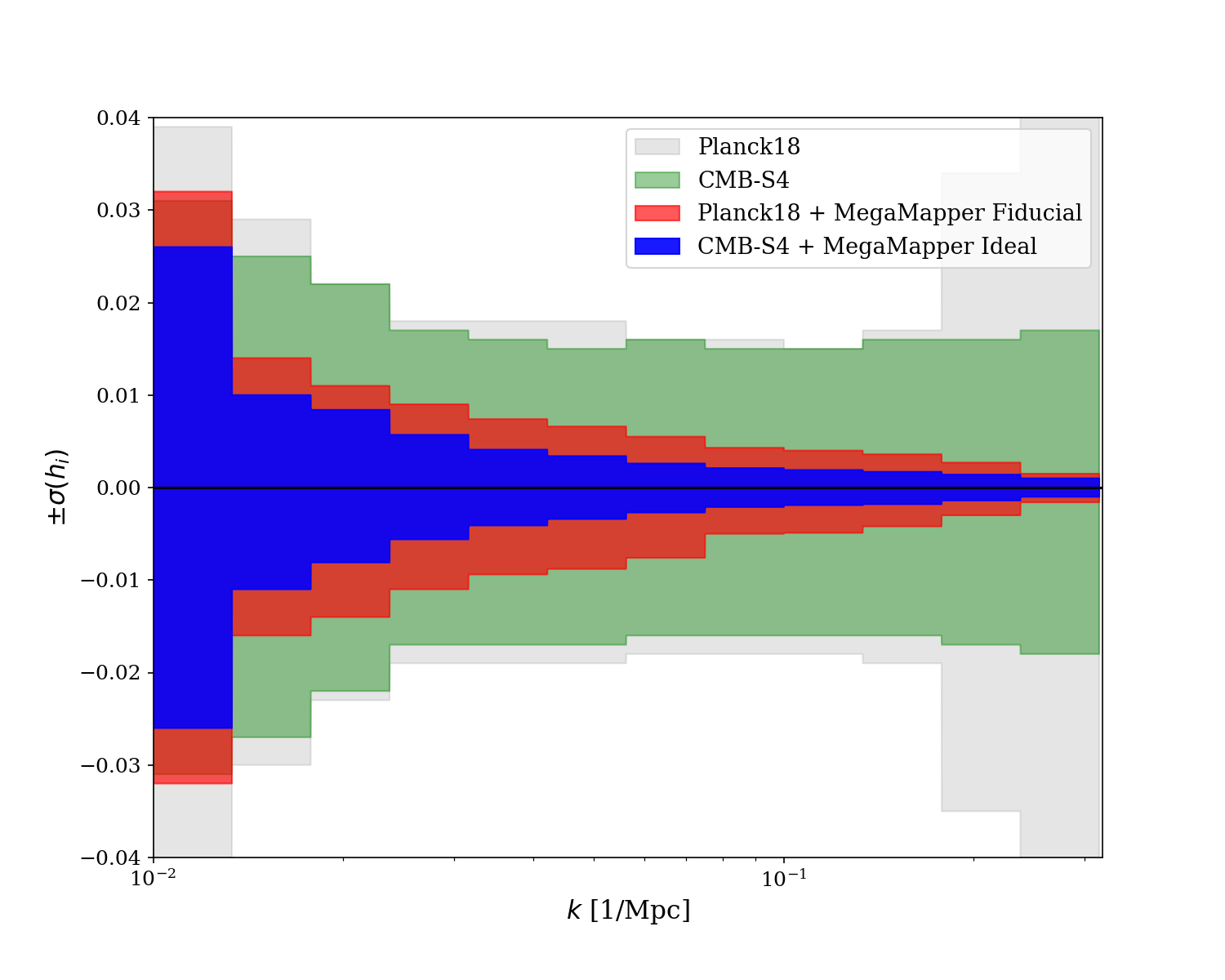}
\caption{Fractional residuals of primordial power spectra residuals relative 
to their best fit power laws, zoomed into 
the $k>0.01\,{\rm Mpc}^{-1}$ range. 
We see each next generation of experiments provides significant improvements in constraining power, with mean precisions over this range of 5.5\% for Planck18 current 
data (grey), 1.8\% for CMB-S4 (green), 0.83\% for Planck18 plus 
MegaMapper Fiducial (red), to 0.57\% for CMB-S4 plus MegaMapper Ideal 
(blue). 
} 
\label{fig:sumppszoom}
\end{figure}

In order to examine more closely the effects of the upcoming data on  
primordial power spectrum reconstruction, and the ability to 
distinguish deviations from a pure power law behavior, 
Fig.~\ref{fig:sumppszoom} shows the fractional residuals of each reconstruction (including CMB-S4 alone) 
relative to its best power law fit, for $k\ge0.01\,{\rm Mpc}^{-1}$ to allow a 
zoomed in scale. The improvements with each successive data set are 
clear.

\subsection{Summary of Neutrino Mass Constraints} 
\label{sec:summnu} 

For neutrino physics in the form of the sum of the neutrino masses 
(and its implications for the neutrino hierarchy), 
large scale structure surveys 
offer great complementarity with CMB experiments. While CMB-S4 
can reduce the uncertainty from the $\sim110\,$meV of Planck 
alone to $\sim50\,$meV, the addition of galaxy redshift survey  
data can give $\sim25\,$meV for Planck plus MegaMapper Fiducial 
or $\sim14\,$meV for CMB-S4 plus MegaMapper Ideal. These constraints 
are not substantially different from those obtained in the pure 
power law case without inflationary freedom, showing that the 
combination of CMB and large scale structure can successfully fit both early and late 
cosmology simultaneously.  

\begin{figure}[htbp!] 
\centering
\includegraphics[width=0.85\columnwidth]{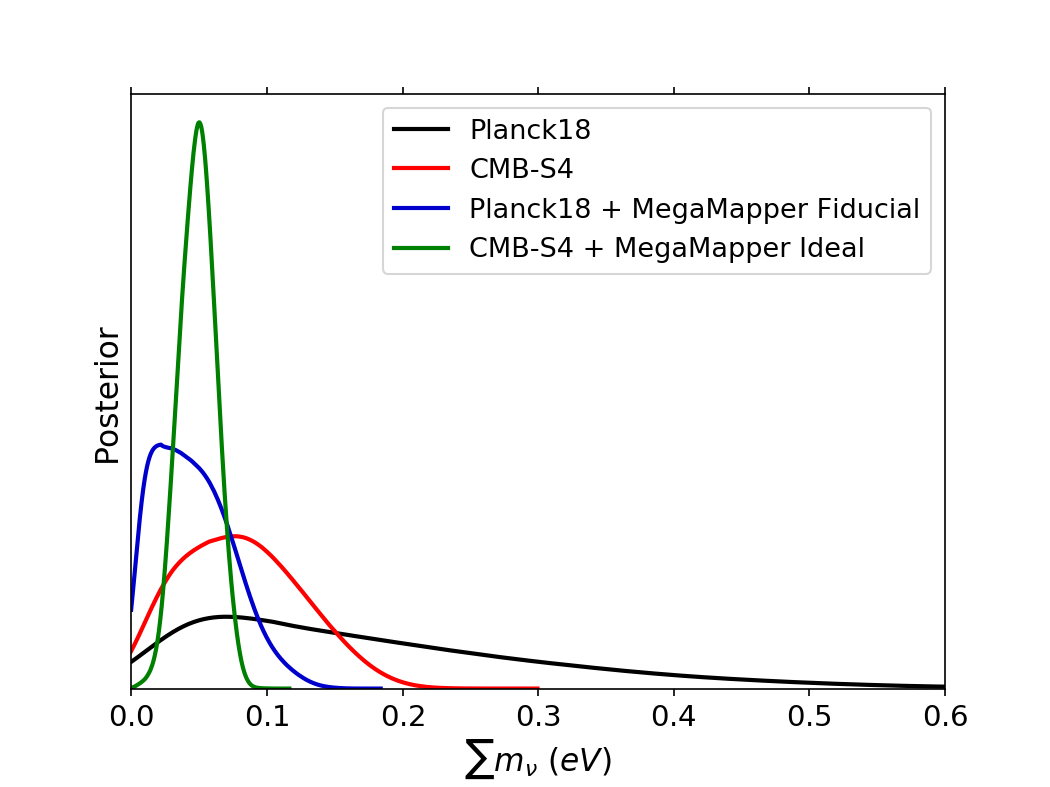}
\caption{
1D PDFs for the sum of neutrino masses are shown for actual data 
from Planck18, and projected for CMB-S4 alone, Planck18 plus MegaMapper 
Fiducial, and CMB-S4 plus MegaMapper Ideal. Each more advanced 
experiment continues improving the constraints, with CMB-S4 plus MegaMapper Ideal capable of clearly ruling out both zero neutrino mass and the inverted hierarchy. 
} 
\label{fig:summnu}
\end{figure}

Figure \ref{fig:summnu} shows the 1D PDF for the sum of neutrino 
masses in these four cases.

\section{Conclusions} \label{sec:concl} 

The very high energy early universe and the physics of inflation 
are outstanding frontiers to explore. Although the simplest 
theories predict a power law form for the primordial power 
spectrum of curvature perturbations, considerable freedom and 
diverse possibilities exist. We allow for inflationary freedom 
and show that sufficiently precise CMB measurements over a range 
of angular scales and large scale structure surveys in the late 
universe over a range of redshifts can not only accommodate such 
freedom but return substantially similar cosmological parameter 
estimation as the restrictive power law assumption. That is, 
future data will allow precise constraints on both the early and 
late universe simultaneously. 

CMB and galaxy redshift surveys work extremely well together, 
and we quantify, by probe or combination of probes, and by 
generation of experiment, how well they can reconstruct the 
primordial power spectrum and late time fundamental parameters 
such as the sum of neutrino masses. 

We reconstruct the primordial power spectrum in a free form 
of 20 bins in wavenumber, allowing stringent tests of the power 
law behavior or revealing features such as steps, bends, or 
oscillations from inflation physics. Due to the power of 
deconvolution of initial conditions and late time transfer 
functions due to the multiple bins in angular scale and redshift 
of the data sets, the free bin amplitudes can be constrained with 
subpercent precisions for $k\gtrsim0.015\,{\rm Mpc}^{-1}$. 
Indeed, some bins can reach 0.1\% precision and for the further 
future combination CMB-S4 plus MegaMapper Ideal the average 
precision over the range $k=0.01$--0.35\,Mpc$^{-1}$ can reach 
0.6\%. 

The 95\% confidence upper limit on the sum of neutrino masses 
-- even allowing for inflationary freedom -- will drop from 
0.45\,eV with current Planck18 data to 0.16\,eV with CMB-S4 
(assuming minimum mass). 
The 68\% confidence level uncertainty will drop from 110 meV 
to 50 meV, and with the addition of MegaMapper Ideal it can 
reach 14 meV. This would allow a stringent test of neutrino 
oscillation bounds on the minimum sum of masses, and possible 
distinction between the normal and inverted hierarchies. An 
advantage of MegaMapper is its high redshift range, $z\gtrsim2$, 
with a huge volume encompassing more modes than measured in 
the CMB, and increased wavenumber range in the linear regime. 
That both makes interpretation of data easier and gives a longer 
lever arm for discernment of scale dependent neutrino mass 
effects. Furthermore we expect the universe to be mostly 
matter dominated there, with reduced dependence on dark energy 
properties. 

If one considers the ``growth'' plane of $\sum\mnu$--$\sigma_8$, 
the progress from Planck18 to CMB-S4 delivers a factor 10 
improvement in the confidence contour area, and adding 
MegaMapper Ideal provides a further factor 10 beyond that. 

Further improvements are possible by including the multipoles 
$\ell<30$ we have left out, either from present data or future 
experiments such as LiteBIRD, and including large scale 
structure surveys at $z<2$. The latter can provide strong 
constraints on dark energy properties beyond \lcdm. 

Overall we have seen that freedom can lead to discovery in the 
early universe, and poses no obstacle to discovery in the 
late universe (i.e.\ the error bars are not appreciably worse than those 
under the power law assumption), with the excellent data sets 
that the next decade will bring.

\section*{Acknowledgments}

This work made use of the CHE cluster, managed and funded by COSMO/CBPF/MCTI, 
with financial support from FINEP and FAPERJ, and operating at the 
Javier Magnin Computing Center/CBPF. This work is supported in part by the Energetic Cosmos Laboratory and by 
the U.S.\ Department of Energy, Office of Science, Office of High Energy 
Physics, under Award DE-SC-0007867 and contract no.\ DE-AC02-05CH11231. 
GB would like to acknowledge the State Scientific and Innovation Funding Agency of Espirito Santo (FAPES, Brazil). 
GB would also like to thank the Brazilian Physical Society (SBF) through the 
SBF/APS PhD Exchange Program for financial support and LBL for the hospitality and 
financial support during the early stages of this work. This
study was financed in part by the Coordena\c{c}\~ao de Aperfei\c{c}oamento de Pessoal de N\'ivel 
Superior -
Brasil (CAPES) - Finance Code 001.

\appendix

\section{More Detailed Results} \label{sec:apxbS}  

Here we give additional information concerning the 
galaxy survey bias and power correction terms, 
and the PPS bin amplitudes, mostly focused on 
covariances. 

\begin{figure}[htbp!] 
\centering
\includegraphics[width=0.95\columnwidth]{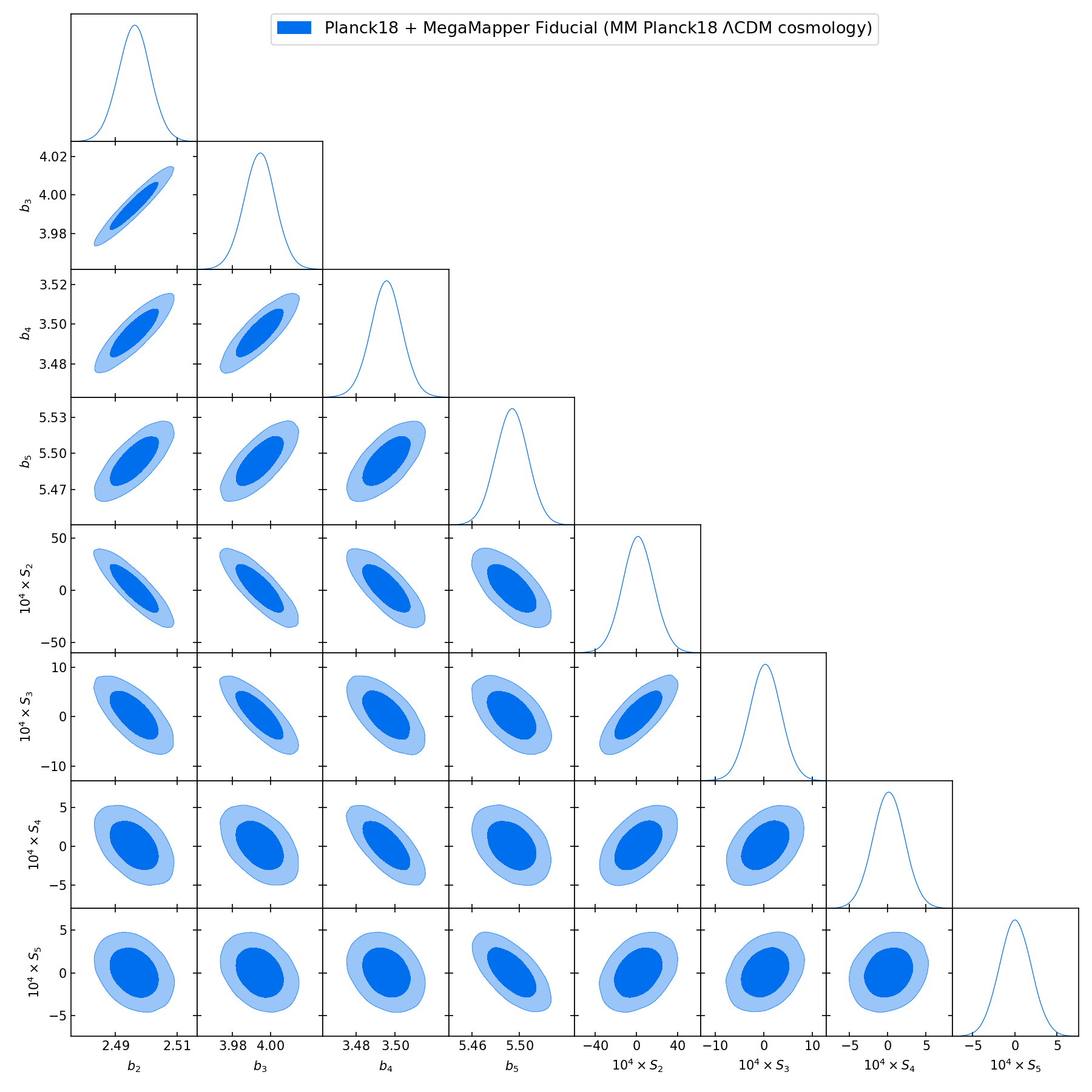}
\caption{
Corner plot of galaxy bias $b_i$ and power correction $S_i$ terms in 
each redshift bin (e.g.\ $b_3$ means the bias in the $z=3$ redshift 
bin), for the case of Planck18 plus MegaMapper Fiducial.  
} 
\label{fig:biSi_18MMf}
\end{figure}

The astrophysics parameters of the galaxy bias $b_i$ and 
power correction $S_i$ in each redshift slice of the  
galaxy redshift survey do not have extreme degeneracies with 
the cosmology parameters discussed in the main text. For 
completeness, we show a corner plot of $b_i$ and $S_i$ against 
each other in Fig.~\ref{fig:biSi_18MMf}, for the case of 
Planck18 plus MegaMapper Fiducial (so four redshift slices). 
This gives a sense for their constraints.

\begin{figure}[htbp!] 
\centering
\includegraphics[width=0.95\columnwidth]{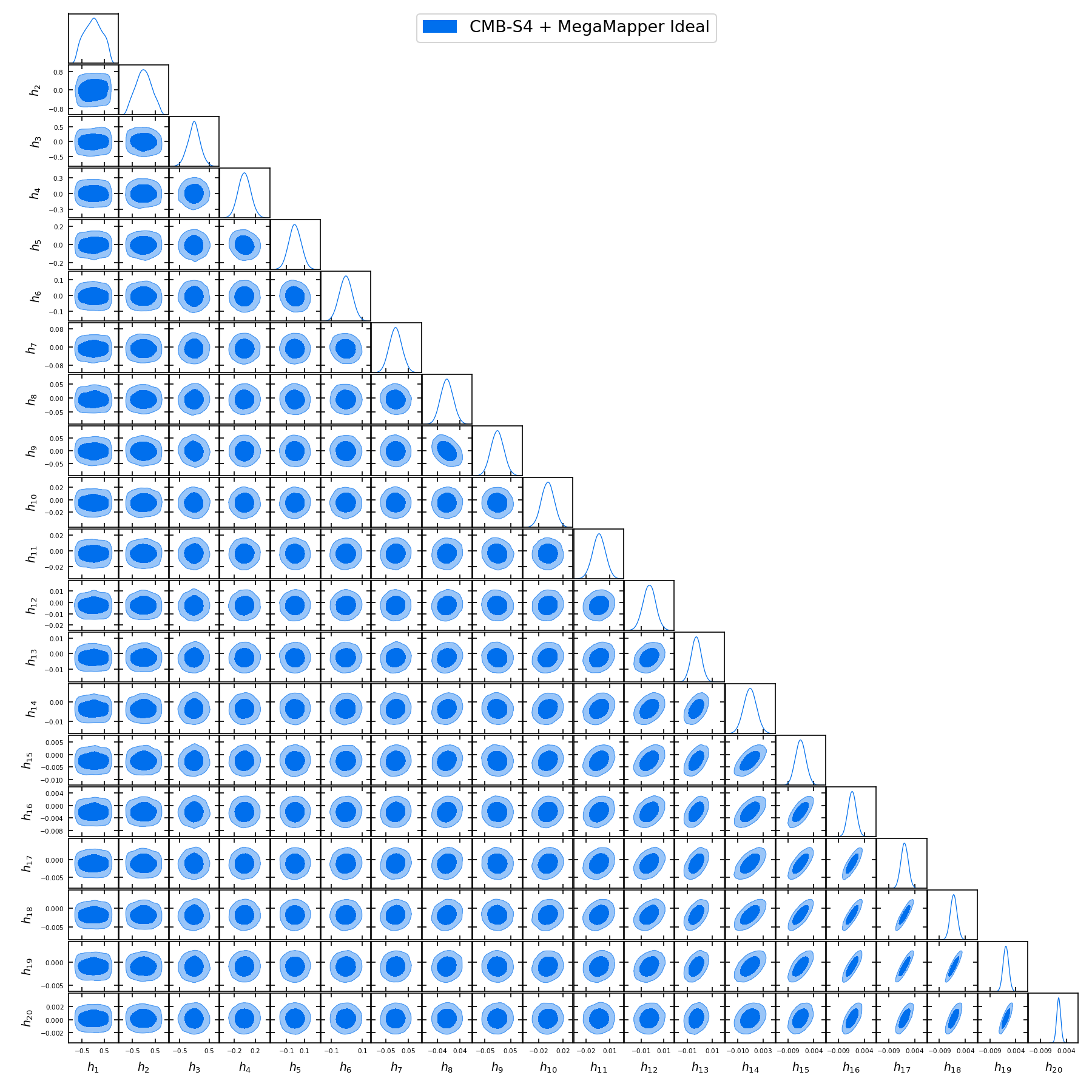}
\caption{
Corner plot of PPS amplitudes $h_i$ in each $k$ bin, for the case of 
CMB-S4 plus MegaMapper Ideal. 
} 
\label{fig:cmbs4MMihi} 
\end{figure}

Similarly, the covariances among the PPS amplitudes 
$h_i$ are moderate, especially with the excellent discrimination 
in wavenumber and redshift of the CMB-S4 plus MegaMapper Ideal surveys. The corner plot is shown in 
Fig.~\ref{fig:cmbs4MMihi}.

\section{Freedom vs Power Law} \label{sec:apxpowerlaw} 

For most of the cosmological parameters the combination of the 
CMB data and galaxy redshift survey data gives strong enough 
constraints over the range of angular scales and redshifts to 
enable comparable parameter fitting to the power law case, despite allowing the extra 
freedom beyond a power law assumption in the PPS. That is, 
the combined data enables exploration of both early and late 
cosmology successfully. 

\begin{figure}[htbp!] 
\centering
\includegraphics[width=0.95\columnwidth]{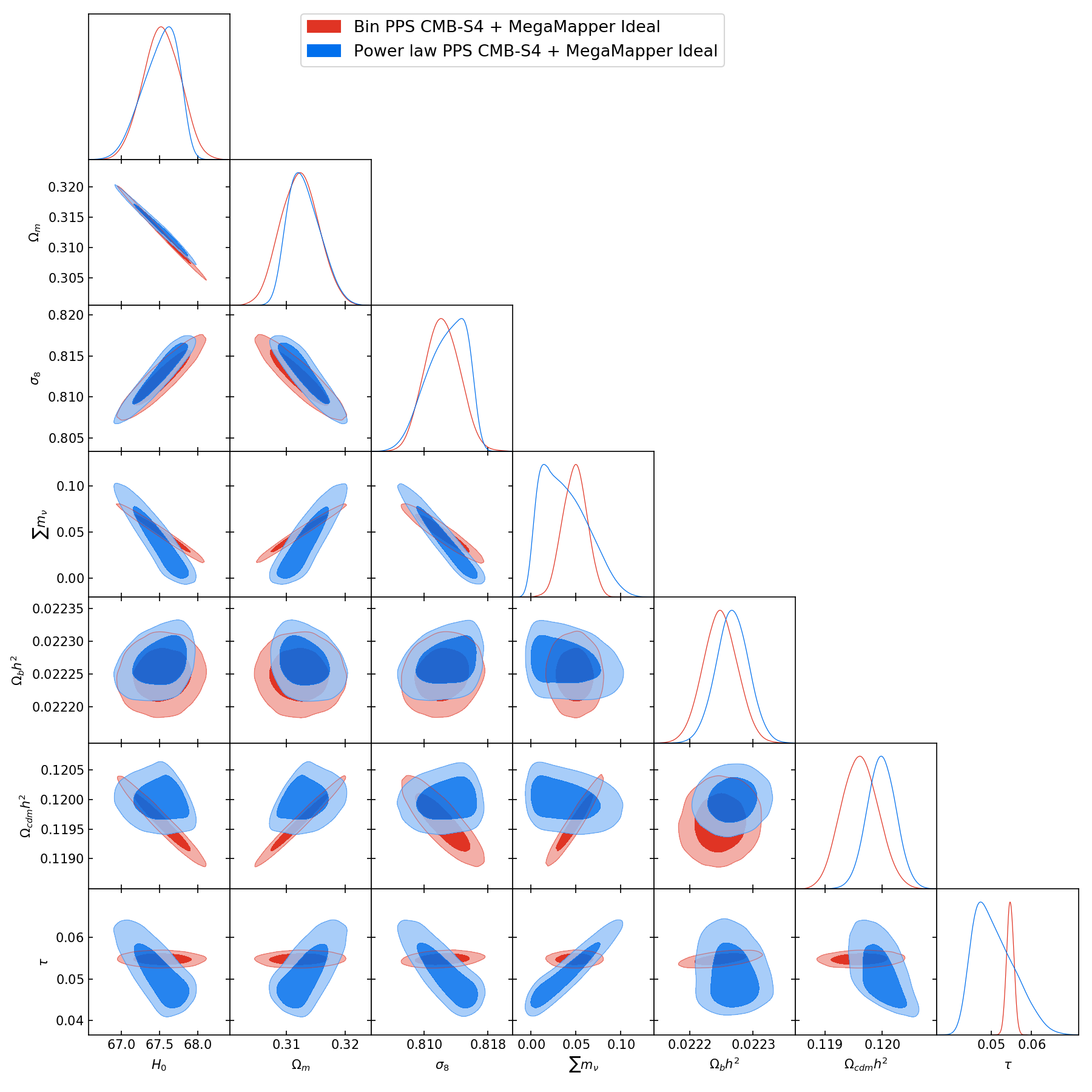}
\caption{
Corner plot of cosmology parameters for CMB-S4 plus MegaMapper Ideal data, comparing fits allowing PPS freedom vs fixing to power law behavior.   
} 
\label{fig:powerlawideal}
\end{figure}

\begin{figure}[htbp!] 
\centering
\includegraphics[width=0.95\columnwidth]{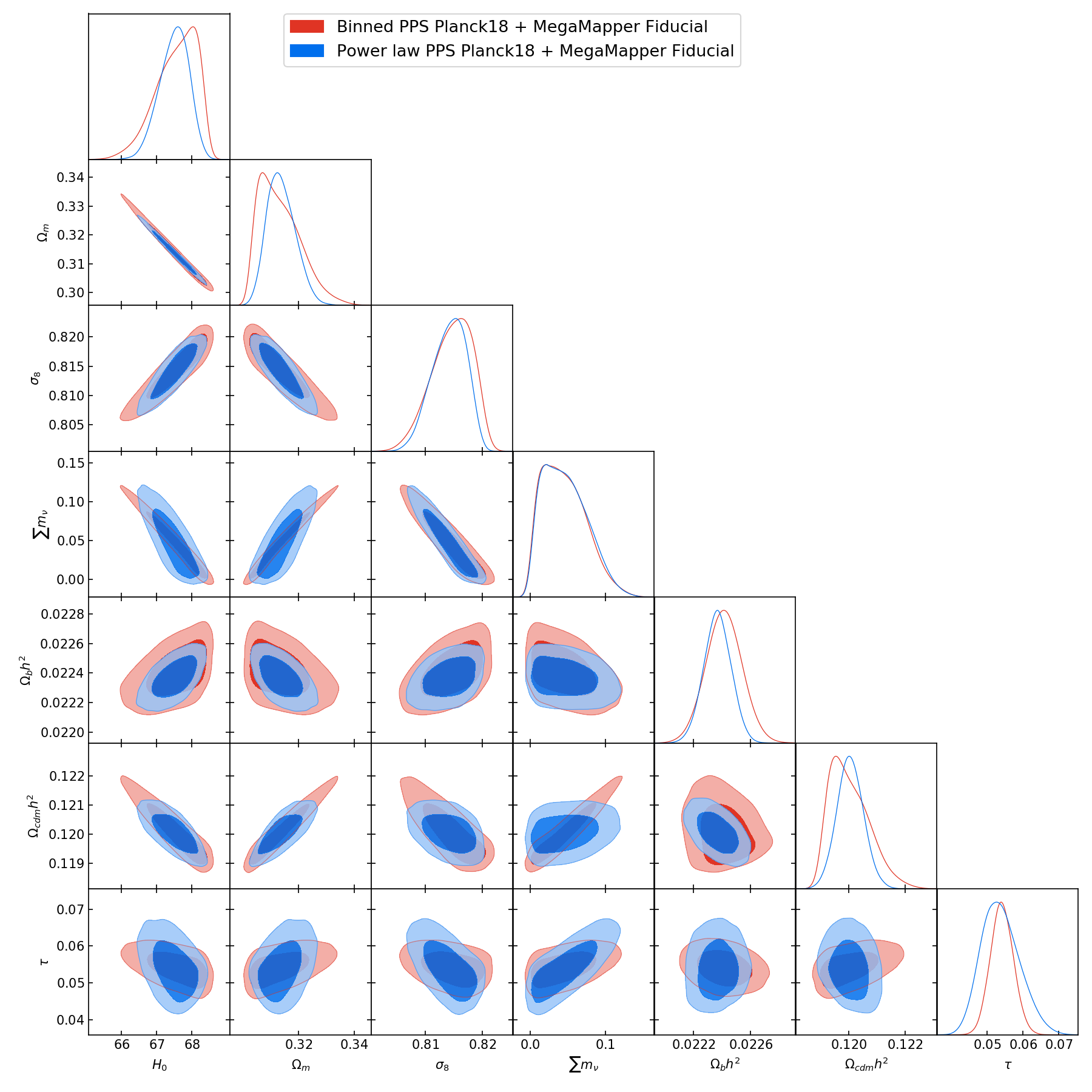}
\caption{
As Fig.~\ref{fig:powerlawideal}, but for Planck18 plus MegaMapper Fiducial. 
} 
\label{fig:powerlawfid}
\end{figure}

Figures~\ref{fig:powerlawideal} and \ref{fig:powerlawfid} 
show the comparison between free and power law PPS results 
for CMB-S4 plus MegaMapper Ideal and Planck18 plus MegaMapper 
Fiducial respectively. Slights offsets and the behavior of 
the optical depth $\tau$ were discussed in 
Sec.~\ref{sec:cmbgal}.

\begin{thebibliography}{99}

\bibitem{deput14}
R. de Putter, E. V. Linder, and A. Mishra, Inflationary Freedom and Cosmological Neutrino
Constraints, Phys. Rev. D 89 (2014), 103502, \href{https://arxiv.org/abs/1401.7022}{arXiv:1401.7022}.

\bibitem{kin00}
W. H. Kinney, How to fool CMB parameter estimation
, Phys. Rev. D 63 (2001), 043001, \href{https://arxiv.org/abs/astro-ph/0005410}{arXiv:astro-ph/0005410}.

\bibitem{haz13}
D. K. Hazra, A. Shafieloo, and T. Souradeep, Cosmological parameter estimation with free-form primordial power spectrum, 
Phys.Rev. D87 (2013), 123528,
\href{https://arxiv.org/abs/1303.5336}{arXiv:1303.5336}.

\bibitem{zhao13}
G.-B. Zhao, S. Saito, W. J. Percival, A. J. Ross, F. Montesano, M. Viel, D. P. Schneider, M. Manera, J. Miralda-Escud\'e, N. Palanque-Delabrouille, et al., 
The clustering of galaxies in the SDSS-III Baryon Oscillation Spectroscopic Survey: weighing the neutrino mass using the galaxy power spectrum of the CMASS sample, 
Mon. Not. Roy. Astron. Soc. 436 (2013), 2038-2053,  
\href{https://arxiv.org/abs/1211.3741}{arXiv:1211.3741}

\bibitem{hann01}
S. Hannestad, 
Reconstructing the inflationary power spectrum from CMBR data, 
Phys. Rev. D 63 (2001), 043009, \href{https://arxiv.org/abs/astro-ph/0009296}{arXiv:astro-ph/0009296}.

\bibitem{hlo11}
R. Hlozek, J. Dunkley, G. Addison, J. W. Appel,
J. R. Bond, C. Sofia Carvalho, S. Das, M. J. Devlin,
R. D\"unner, T. Essinger-Hileman, et al., 
The Atacama Cosmology Telescope: a measurement of the primordial power spectrum, 
Astrophys. J. 749 (2012) 90, \href{https://arxiv.org/abs/1105.4887}{arXiv:1105.4887}.

\bibitem{gau12}
C. Gauthier and M. Bucher, Reconstructing the primordial power spectrum from the CMB, JCAP 1210 (2012) 050, 
\href{https://arxiv.org/abs/1209.2147}{arXiv:1209.2147}.

\bibitem{dival16}
E. Di Valentino, S. Gariazzo, M. Gerbino, E. Giusarma,
and O. Mena, 
Dark Radiation and Inflationary Freedom after Planck 2015, 
Phys. Rev. D 93 (2016), 083523,
\href{https://arxiv.org/abs/1601.07557}{arXiv:1601.07557}.

\bibitem{gar15}
S. Gariazzo, L. Lopez-Honorez, and O. Mena, Primordial power spectrum features and $f_{\rm NL}$ constraints, Phys. Rev.
D 92 (2015), 063510, \href{https://arxiv.org/abs/1506.05251}{arXiv:1506.05251}.

\bibitem{rav16}
A. Ravenni, L. Verde, and A. J. Cuesta, Red, Straight, no bends: primordial power spectrum
reconstruction from CMB and large-scale structure, JCAP 1608 (2016) 028, \href{https://arxiv.org/abs/1605.06637}{arXiv:1605.06637}.

\bibitem{can16}
N. Canac, G. Aslanyan, K. N. Abazajian, R. Easther, and L. C. Price, Testing for New Physics: Neutrinos and the Primordial Power Spectrum, JCAP 1609 (2016) 022, 
\href{https://arxiv.org/abs/1606.03057}{arXiv:1606.03057}.

\bibitem{hunt15}
P. Hunt and S. Sarkar, Search for features in the spectrum of primordial perturbations
using Planck and other datasets, JCAP 1512 (2015) 052,
\href{https://arxiv.org/abs/1510.03338}{arXiv:1510.03338}.

\bibitem{haz18}
D. K. Hazra, D. Paoletti, M. Ballardini, F. Finelli,
A. Shafieloo, G. F. Smoot, and A. A. Starobinsky, Probing features in inflaton potential and reionization history with future CMB space observations, JCAP 1802 (2018) 017,
\href{https://arxiv.org/abs/1710.01205}{arXiv:1710.01205}.

\bibitem{haz19}
 D. K. Hazra, A. Shafieloo, and T. Souradeep, Parameter discordance in Planck CMB and low redshift measurements: projection in the primordial power spectrum, JCAP 1904 (2019) 036,
\href{http://arxiv.org/abs/1810.08101}{arXiv:1810.08101}.

\bibitem{sloz19}
A. Slosar, X. Chen, C. Dvorkin, D. Green, P. D. Meerburg, E. Silverstein et al., Scratches
from the Past: Inflationary Archaeology through Features in the Power Spectrum of
Primordial Fluctuations, \href{https://arxiv.org/abs/1903.09883}{arXiv:1903.09883}.

\bibitem{chluba15}
J. Chluba, J. Hamann and S. P. Patil, Features and New Physical Scales in Primordial Observables: Theory and Observation, Int. J. Mod. Phys. D 24 (2015), 1530023, \href{https://arxiv.org/abs/1505.01834}{arXiv:1505.01834}.

\bibitem{beutl19}
F. Beutler, M. Biagetti, D. Green, A. Slosar, B. Wallisch, Primordial Features from Linear to Nonlinear Scales, 
Phys. Rev. Research 1 (2019) 033209, \href{https://arxiv.org/abs/1906.08758}{arXiv:1906.08758}.

\bibitem{zeng18}
C. Zeng, E. Kovetz, X. Chen, J. Munoz, and M. Kamionkowski, Searching for Oscillations in the
Primordial Power Spectrum with CMB and LSS Data, Phys. Rev. D 99 (2019), 043517,
\href{https://arxiv.org/abs/1812.05105}{arXiv:1812.05105}.

\bibitem{domn19}
G. Domnech and M. Kamionkowski, Lensing anomaly and oscillations in the primordial power spectrum, JCAP 1911 (2019) 040 \href{https://arxiv.org/abs/1905.04323}{arXiv:1905.04323}.

\bibitem{handley19}
W. J. Handley, A. N. Lasenby, S. P. Patil, V. H. Peiris and M. P. Hobson, Bayesian inflationary reconstructions from Planck 2018 data, Phys. Rev. D 100 (2019), 103511, \href{https://arxiv.org/abs/1908.00906}{arXiv:1908.00906}.

\bibitem{lecl19}
F. Leclercq, W. Enzi, J. Jasche, A. Heavens, Primordial power spectrum and cosmology from black-box galaxy surveys, Mon. Not. Roy. Astron. Soc. 490 (2019), 4237, \href{https://arxiv.org/abs/1902.10149}{arXiv:1902.10149}.

\bibitem{sobs}
 Simons Observatory Collaboration, The Simons Observatory: Science
goals and forecasts, JCAP 1902 (2019) 056, \href{https://arxiv.org/abs/1808.07445}{arXiv:1808.07445}.

\bibitem{plc18inflation}
Planck Collaboration, Planck 2018 results. X. Constraints on inflation, 
\href{https://arxiv.org/abs/1807.06211}{arXiv:1807.06211}.

\bibitem{chen16}
X. Chen, C. Dvorkin, Z. Huang, M. H. Namjoo, and L. Verde, The Future of
Primordial Features with Large-Scale Structure Surveys, JCAP 1611 (2016) 014, \href{https://arxiv.org/abs/1605.09365}{arXiv:1605.09365}.

\bibitem{haz16}
D. K. Hazra, A. Shafieloo, G. F. Smoot and A. A. Starobinsky, Primordial features and Planck polarization, 
JCAP 1609 (2016) 009, \href{https://arxiv.org/abs/1605.02106}{arXiv:1605.02106}.

\bibitem{lhui17}
B. L'Huillier, A. Shafieloo, D. K. Hazra, G. Smoot, and A. Starobinsky, Probing Features in the
Primordial Perturbation Spectrum with Large-Scale Structure Data, Mon. Not. Roy. Astron. Soc. 477 (2018), 2503,
\href{https://arxiv.org/abs/1710.10987}{arXiv:1710.10987}. 

\bibitem{obied17}
G. Obied, C. Dvorkin, C. Heinrich, W. Hu, and V. Miranda, Inflationary Features and Shifts in Cosmological Parameters from Planck 2015 Data, 
Phys. Rev. D 96 (2017), 083526, \href{https://arxiv.org/abs/1706.09412}{arXiv:1706.09412}. 

\bibitem{planck2018}
Planck Collaboration, Planck 2018 results. VI. Cosmological parameters, \href{https://arxiv.org/abs/1807.06209}{arXiv:1807.06209}.

\bibitem{class}
D. Blas, J. Lesgourgues, T. Tram, CLASS II: Approximation schemes, 
JCAP 1107 (2011) 034, \href{https://arxiv.org/abs/}{arXiv:1104.2933}.

\bibitem{jesus}
A. Ach\'ucarro, V. Atal, P. Ortiz, and J. Torrado, Localized correlated features in the CMB power spectrum and primordial bispectrum from a transient reduction in the speed of sound, Phys. Rev. D 89 (2014), 103006, \href{https://arxiv.org/abs/1311.2552}{arXiv:1311.2552}.

\bibitem{cyr13}
 F.-Y. Cyr-Racine, R. de Putter, A. Raccanelli and K. Sigurdson, Constraints on Large-Scale
Dark Acoustic Oscillations from Cosmology, Phys. Rev. D 89 (2014), 063517, \href{https://arxiv.org/abs/1310.3278}{arXiv:1310.3278}.

\bibitem{cmbs4sci}
CMB-S4 collaboration, CMB-S4 Science Book, First Edition,  \href{https://arxiv.org/abs/1610.02743}{arXiv:1610.02743}.

\bibitem{cmbs4des}
K. Abazajian et al., CMB-S4 Science Case, Reference Design, and Project Plan, \href{https://arxiv.org/abs/1907.04473}{arXiv:1907.04473}.

\bibitem{MM1}
D.J. Schlegel et al., Astro2020 APC white paper: the MegaMapper: a $z > 2$ spectroscopic
instrument for the study of inflation and dark energy, \href{https://arxiv.org/abs/1907.11171}{arXiv:1907.11171}.

\bibitem{MM2}
S. Ferraro et al., Inflation and Dark Energy from spectroscopy at $z > 2$,
\href{https://arxiv.org/abs/1903.09208}{arXiv:1903.09208}.

\bibitem{montepython1}
B. Audren, J. Lesgourgues, K. Benabed, and S. Prunet,
Conservative constraints on early cosmology: An illustration of the Monte Python cosmological parameter inference code, JCAP 1302 (2013) 001, \href{https://arxiv.org/abs/1210.7183}{arXiv:1210.7183}.

\bibitem{montepython2}
T. Brinckmann and J. Lesgourgues, MontePython 3:
Boosted MCMC sampler and other features, Phys. Rev.
D 97 (2018), 063506, \href{https://arxiv.org/abs/1804.07261}{arXiv:1804.07261}.

\bibitem{spreng18}
T. Brinckmann, D. C. Hooper, M. Archidiacono, J. Lesgourgues, and T. Sprenger, The
promising future of a robust cosmological neutrino mass measurement, JCAP 1901 (2019)
059, \href{https://arxiv.org/abs/1808.05955}{arXiv:1808.05955}.

\bibitem{getdist}
A. Lewis, GetDist: a Python package for analysing
Monte Carlo samples, \href{https://arxiv.org/abs/1910.13970}{arXiv:1910.13970}, \href{https://getdist.readthedocs.io/en/latest/}{https://getdist.readthedocs.io}.

\bibitem{litebird1}
T. Matsumura et al., Mission design of LiteBIRD, Journal of 
Low Temperature Physics, 176 (2014) 733, \href{https://arxiv.org/abs/1311.2847}{arXiv:1311.2847}.

\bibitem{litebird2} 
H. Sugai et al., Updated design of the CMB polarization experiment satellite LiteBIRD, Journal of Low Temperature 
Physics (2020), in press, \href{https://arxiv.org/abs/2001.01724}{arXiv:2001.01724}. 


\end{thebibliography}
\end{document}